# Experimental Clocking of Nanomagnets with Strain for Ultra Low Power Boolean Logic

Noel D'Souza[a], Mohammad Salehi Fashami[a], Supriyo Bandyopadhyay[b] and Jayasimha Atulasimha[a,*]

[a]Department of Mechanical and Nuclear Engineering

[b]Department of Electrical and Computer Engineering

Virginia Commonwealth University, Richmond, VA 23284, USA

**Nanomagnetic implementations of Boolean logic[1,2] have garnered attention because of their non-volatility and the potential for unprecedented energy-efficiency. Unfortunately, the large dissipative losses that occur when nanomagnets are switched with a magnetic field[3] or spin-transfer-torque[4] inhibit the promised energy-efficiency. Recently, there have been experimental reports of utilizing the Spin Hall effect for switching magnets[5–7], and theoretical proposals for strain induced switching of single-domain magnetostrictive nanomagnets[8–12], that might reduce the dissipative losses significantly. Here, we experimentally demonstrate, for the first time, that strain-induced switching of single-domain magnetostrictive nanomagnets of lateral dimensions ~200 nm fabricated on a piezoelectric substrate can implement a nanomagnetic Boolean NOT gate and unidirectional bit information propagation in dipole-coupled nanomagnet chains. This portends ultra-low-energy logic processors and mobile electronics that may operate solely by harvesting energy from the environment without ever needing a battery.**

* Correspondence to be addressed to jatulasimha@vcu.edu

Nanomagnet-based logic switches[1,2], in which logic bits 0 and 1 are encoded in two stable magnetization orientations along the easy (major) axis of a shape-anisotropic elliptical single-domain nanomagnet, and in which switching is accomplished by flipping the magnetization from one stable orientation to the other, have emerged as potential replacements for current complementary metal-oxide-semiconductor (CMOS) transistor switches because of superior energy-efficiency. In this letter, we demonstrate strain-induced switching of nanomagnets that could render nanomagnetic logic 2-3 orders of magnitude more energy-efficient than conventional transistor-based logic. A transistor dissipates at least ~$10^4$ kT of energy to switch in isolation[13] and $10^5$ kT to switch in a circuit in a reasonable time of ~1 ns. In contrast, a magnetic binary switch may dissipate a mere ~$10^2$ kT of energy to switch in ~1 ns if implemented with an elliptical, two-phase composite multiferroic nanomagnet consisting of a single-domain magnetostrictive layer elastically coupled to an underlying piezoelectric layer[8–10]. When a tiny electrostatic potential is applied across the piezoelectric layer, it deforms and the resulting strain is transferred to the magnetostrictive layer, making its magnetization rotate by a large angle as shown in Fig. 1. Such rotations can be utilized to write bits in non-volatile memory[12,14,15] or implement Bennett-clocked logic gates in the fashion of magnetic quantum cellular automata[1,2,8,10]. So far, several experimental studies have been performed to demonstrate strain-induced magnetization switching[16–19], but only in magnets that are either multi-domain[20], or where strain moves domain walls to switch the magnetization (instead of rotating it)[21–23] or in single-domain nanomagnets where the coherent rotation is ~90º and not ~180º[24]. The experimental studies in this work demonstrate, for the first time, strain-induced 180º switching of single-domain magnetostrictive nanomagnets on a piezoelectric substrate to realize a Boolean NOT logic gate and unidirectional propagation of logic bit information down a chain of nanomagnets. These are the key steps in the realization of strain-clocked nanomagnetic logic and information processing.

Strain-induced switching of magnetization is demonstrated using magnetostrictive Co nanomagnets (nominal diameter ~200 nm) deposited on a (001) Pb(Mg$_{1/3}$Nb$_{2/3}$)O$_3$–PbTiO$_3$ (PMN-PT) 70/30 substrate of dimensions 5×5×0.5 mm$^3$. The experimental setup is described in Supplementary section A. At first, the substrate is poled with an 800 kV m$^{-1}$ electric field and subsequently a linear strain-field characteristic is observed up to 400 kV m$^{-1}$. Next, the magnets are deposited on the poled substrate and their magnetizations "initialized" to the 'down' direction (↓) by applying a magnetic field of ~200 mT along the easy axis of the nanomagnets. Finally, an electric field of 400 kV m$^{-1}$ is applied along the poling direction that generates a strain of ~400 ppm in PMN-PT which is mostly transferred to the ~12 nm thick nanomagnets and produces a stress of ~80 MPa therein (cobalt's Young's Modulus ~200 GPa[25]). The results of the material characterization of the cobalt thin-film (SEM, EDS, M-H curves, etc.) and the PMN-PT substrate (surface roughness, EDS) are shown in Supplementary Section B. The negligible effect of the thin CoO layer (< 2 nm over a period of several weeks) that develops on the Co nanomagnets owing to surface oxidation is also discussed in Supplementary section B(c). Three different cases are discussed below:

CASE I: Isolated nanomagnets (Fig. 1): We study an array of nanomagnets with inter-magnet spacing (~800 nm) large enough to disallow any significant dipole interaction between neighbours. They are all initially magnetized in the same ("down") direction as shown in the left panel of Fig. 1a and the right panel indicates the direction in which we expect the magnetization to rotate when a tensile (+$\sigma$) or compressive (-$\sigma$) stress is applied. Figs. 1b and 1c show experimental MFM images of the nanomagnets prior to and after application of tensile stress.

In Fig. 1b (magnet volume nominally 250×150×12 nm$^3$), the stress generated is not large enough to overcome the shape anisotropy of the magnets and make the magnetizations rotate. Thus, the pre-stress (Fig. 1b, left) and post-stress (Fig. 1b, right) magnetic states are identical. Fig. 1c (magnet volume nominally 200×175×12 nm$^3$) shows that nanomagnets with lower shape anisotropy do experience magnetization rotation. When stress is applied to these nanomagnets, their magnetizations orient themselves along the hard axis by rotating through ~90º. Upon removal of the stress, the magnetizations have equal probability of returning to their initial orientations or flipping to the opposite directions. Hence, one would expect 50% of the magnets will flip their magnetization orientation from "down" to "up" and the rest would flip back to the original "down" state. However, owing to uncontrollable factors such as lithographic variances, surface roughness, stress concentration, etc., only a fraction of the magnets meet the correct condition (stress anisotropy greater than shape anisotropy to allow ~90º rotation), resulting in far fewer than 50% of the magnets flipping their magnetization orientations by 180º. The magnet dimensions are chosen to ensure that the shape anisotropy is high enough to allow good MFM imaging and yet small enough to allow stress to rotate the magnetization (see supplementary section A).

CASE II: Two dipole-coupled nanomagnets: Boolean NOT gate (Fig. 2): Consider two elliptical nanomagnets as shown in Fig. 2a that are spaced close enough to allow significant dipole coupling. Since the line joining their centres lie along the minor axes of the ellipses, the dipole coupling will favour anti-parallel ordering. Each nanomagnet encodes a logic bit in its magnetization orientation (say, the "up" orientation encodes bit 1 and "down" orientation bit 0). The magnetization orientation of the left and right magnets represents the input and output bit respectively. The anti-parallel ordering should make the output bit the logic complement of the input bit and make the magnet pair act as a NOT gate, but this is not automatic. Suppose that the right magnet's orientation was initially "down" and an input bit ("0") arrived to orient the left magnet's orientation to the "down" state, thereby leaving both the magnets in the "down" state denoted by (↓↓) as in Fig. 2a(i). While the dipole coupling prefers the (↓↑) state, it is not strong enough (centre-centre distance ~300 nm) to make the right magnet's (R) magnetization overcome its own shape anisotropy energy barrier and flip to assume the "up" orientation. To make it do so, we need to "clock" the magnetostrictive nanomagnets with stress. The left magnet is deliberately designed to be more shape anisotropic (~250×150×12 $nm^3$) than the right (~200×175×12 $nm^3$). Therefore, a global strain/stress that affects both nanomagnets will rotate only the right nanomagnet's magnetization if the stress is strong enough to overcome the shape anisotropy of the right but not the left nanomagnet. This ensures unidirectionality in information propagation, i.e. the magnetization state of the left influences that of the right, but not vice versa.

Thus, the system reaches the (↓→) state. We explain in Supplementary Fig. S19 how this unidirectionality can be achieved in identical nanomagnets having the same shape anisotropy.

Upon removal of the stress (voltage), the magnetization of the output magnet (R) will prefer to assume the "up" orientation over the "down" orientation because of the dipole interaction with its left neighbour. It will therefore flip "up" with very high probability and implement the NOT function by reaching the (↓↑) configuration. The stress has acted as a "Bennett clock"[26] to remove the potential barrier between the local (↓↓) and global minima (↓↑), thereby enabling the right magnet's magnetization to migrate from the local to the global minimum and implementing the NOT operation as explained by the energy profiles in Fig. 2a (bottom panel). Had we started with an (↑↑) configuration and applied the above stress "clock", we would have reached the (↑↓) state. Thus, the "NOT" operation works for either input bit.

The experimental results are shown in Fig. 2b. When ~80 MPa of stress is applied to the bulk substrate, the magnetization of the "input" magnet (L) does not rotate significantly, while that of the "output" magnet (R) rotates by ~90°. When stress is removed, the magnetization of magnet (R) flips "up" due to its dipole interaction with magnet (L). Thus, the magnetization state of this dipole pair changes from its pre-stress state of (↓↓) to a post-stress state of (↓↑), as highlighted by the yellow arrow in Fig. 2b, implementing a logical NOT operation. As explained earlier and discussed in Supplementary Section C(b), because of lithographic variances, all "output" magnets (R) do not flip; only a small fraction do. This is not a shortcoming of the scheme; rather, it is due to our inability to carry out flawless nanolithography.

CASE III: An array of three dipole-coupled magnets (Fig. 3):

Strain clocking can implement a *unidirectional* "binary wire" that propagates a logic bit unidirectional along an array of three nanomagnets (Fig. 3a) of decreasing shape anisotropy with magnet L (~250×150×12 nm$^3$) > magnet C (~200×175×12 nm$^3$) > magnet R (~200×185×12 nm$^3$), each separated by 300 nm (centre-centre distance). Again, we assume that we initialize the array with a global magnetic field so that the magnetizations of all the nanomagnets point "down", as represented by the state (↓↓↓). Upon application of a stress sufficient to overcome the shape anisotropy barriers of magnets (C) and (R), their magnetizations rotate to align along the hard axis, while the magnetization of (L) shows little or no rotation. This takes the system to the (↓→→) state. As the stress is gradually withdrawn, the system passes through an intermediate stage where the shape anisotropy of magnet (C) begins to exceed the stress anisotropy. At this stage, the dipole interaction with (L) forces the magnet (C) to rotate to the "up" state. Note that the stress is still high enough to ensure that the magnet (R), with the weakest shape anisotropy, still points along the hard axis and the system is in the (↓↑→) state. Finally, as the stress is reduced further (and eventually removed) the magnetization of magnet (R) rotates under the dipole influence of the magnet (C) and the system settles to a (↓↑↓) state. We can view this as the bit information encoded in magnet (L) having propagated unidirectionally through magnet (C) to magnet (R).

The MFM results of this case are shown in Fig. 3b. When a stress of ~80 MPa is applied to the bulk substrate and gradually withdrawn, the magnetic state (↓↑↓) as shown in Fig. 3b (yellow arrows) is achieved. This demonstrates logic bit propagation down a chain of nanomagnets in a unidirectional manner, resulting in a unidirectional logic "wire" that is needed to ferry logic bits from one stage to another. Only the initial and final magnetic states are shown since the MFM cannot be performed while voltage/stress is applied. We also note that fabrication defects result in some nanomagnets not switching at all or switching to wrong states. This is discussed in detail in Supplementary Section C.a.3.

These results, combined with the MFM images of Supplementary Section C ("initialized" to (↑)), demonstrate clocking of nanomagnets in both directions, from (↓) to (↑) and from (↑) to (↓), as well their corresponding dipole-coupled scenarios implementing basic NOT logic functionality and information propagation.

We discussed the reasons for the low yield (small fractions of nanomagnets switching) in Supplementary section C. The main issue is lithographic variations. We also discuss the issue of repeatability (the number of times the same nanomagnet switches on repeating the experiment) and explain why materials with high magnetoelastic coupling (such as Terfenol-D) may be needed to improve the switching statistics.

We have demonstrated nanomagnetic logic by applying a *global* stress using a bulk substrate. While this does not directly demonstrate energy-efficiency due to the large voltages applied, these experiments lay the foundation for future energy-efficient Boolean computation devices using this strain based clocking. Ultimately, one would like to implement these operations in chains of nominally identical nanomagnets with a local stress applied in a phased manner to ensure unidirectional propagation of information. Local stress is generated in a targeted nanomagnet in the manner of Cui et al.[19] and is elucidated in Supplementary Fig. S19. In the Supplementary Section D, we also estimate that the energy dissipated per clock cycle in the local scheme (for appropriate materials and dimensions) is ~ 1 aJ/bit and can potentially be an order of magnitude smaller than that estimated for switching magnets with the spin Hall effect[6]. This makes strain based clocking the most energy efficient nanomagnetic clocking mechanism extant.

In summary, we have demonstrated strain-clocked nanomagnetic logic utilizing single-domain Co nanomagnets of ~200 nm lateral dimensions on a bulk PMN-PT substrate. The miniscule energy that will be dissipated per bit flip (~1 aJ for appropriate materials and dimensions) could enable low-density processors, with ~$10^6$ switches and experiencing ~10% activity level (i.e. 10% of the switches flipping at any given time), and clocked at 1 GHz, to dissipate only 100 μW. Such small power requirements can be met by harvesting energy from the surroundings (vibration, TV networks, 3G, etc.) without requiring a battery[27,28]. Provided lithographic imperfections and noise-induced switching errors[29,30] can be mitigated, and nanomagnets with higher magnetoelastic coupling (e.g. Terfenol-D) can be successfully fabricated, strain-clocked nanomagnetic logic processors could lead to a new genre of devices with unprecedented potential. They could transform medically implanted processors that monitor vital body functions[31], human-powered wearable computers[32] and sensors embedded in structures (tall buildings, bridges) that continuously monitor fracture, material fatigue, etc[33] as these applications benefit immensely by eliminating the need for a power source.

**Methods Summary**

A bulk (001) PMN-PT 70/30 substrate of dimensions (5×5×0.5) mm$^3$ was initially poled along the length with an electric field of 800 kV m$^{-1}$ in a castor oil bath. Subsequently, the substrate was cleaned in acetone and IPA and a bilayer of positive e-beam resist (495K PMMA and 950K PMMA; 2% Anisole) was spin-coated as follows:

A static dispense of ~3 ml (495K PMMA) was carried out on the PMN-PT substrate followed by a dynamic spread at 500 rpm for 5 seconds. The spin cycle was performed at a rate of 4000 rpm for 45 seconds. A pre-bake at 115°C (so as not to exceed the PMN-PT Curie temperature of 150°C) was then performed for 90 seconds, resulting in a 495K PMMA layer of ~50 nm. The top 950K PMMA layer was spin-coated next using the same procedure.

Electron-beam lithography was performed at 30 kV using a Hitachi SU-70 SEM in conjunction with the Nabity NPGS nanolithography system. A beam current of 60 pA and dose of 150 – 250 μC cm$^{-2}$ was used to create the elliptical structures. The PMMA-coated substrate is then developed in an MIBK:IPA (1:3) [(methyl isobutyl ketone: isopropyl alcohol)] solution for 70 seconds, rinsed in IPA for 20 seconds to remove the exposed PMMA and finally blow-dried.
A 12 nm layer of Co (above a 5 nm Ti adhesion layer deposited at 0.5 angstrom/s) was then deposited at 0.3 angstrom/s using an electron-beam evaporator at a base pressure of ~3.5 × 10$^{-7}$ Torr. Finally, lift-off was performed by soaking the substrate in acetone for ~5 minutes at 30° C and using an ultrasonic cleaner for 10 seconds to strip off the Ti/Co layers above the unexposed PMMA regions. Magnetic characterization of the elliptical nanomagnets is performed using a Veeco Atomic Force Microscope (AFM) with low-moment magnetic force microscope tips (Bruker MESP-LM) at a lift height of 60 nm.

**Supplementary Information** is linked to the online version of the paper at www.nature.com/nnano

**Acknowledgements** All authors acknowledge support from the National Science Foundation (NSF) under the NEB2020 grant ECCS-1124714 and SHF Grant CCF-1216614 as well as the Semiconductor Research Company (SRC) under NRI task 2203.001. J. A. and N.D. also acknowledge support from NSF CAREER grant CCF-1253370.

**Author Contributions**
All authors played a role in conception of the idea, planning the experiments, discussing the data, analysing the results and writing the manuscript. N.D. fabricated the devices and characterized the nanomagnetic switching. M.S.F. assisted with the fabrication and Co deposition.

**Additional Information** Reprints and permissions information is available at www.nature.com/reprints. The authors declare no competing financial interests. Correspondence and requests for materials should be addressed to J.A. (jatulasimha@vcu.edu).

**Competing Financial Interests**

The authors declare no competing financial interests.

# Figure Captions

**Figure 1: Clocking of single-domain magnetostrictive nanomagnets on bulk piezoelectric substrate and MFM phase images of Co nanomagnets on PMN-PT substrate in pre- and post-stress states. a,** (Left panel) Elliptical nanomagnets initially magnetized along one of their two stable orientations. (Right panel) Electric field applied to bulk (001) PMN-PT substrate along the length results in a stress, $\pm \sigma$, in the direction of the field via $d_{33}$ coupling, which is transferred to the Co nanomagnets. A compressive stress results in the magnetization of the nanomagnet aligning itself along the direction of stress application, while a tensile stress rotates the magnetization to a direction perpendicular the stress axis. We show the stress producing a coherent 90 degree rotation for the sake of illustration and ease of understanding. **b,** Nanomagnets of nominal dimension (250×150×12 nm) having a shape anisotropy energy much higher than that of stress anisotropy energy at ~80 MPa stress. As a result, the magnetization orientations of the nanomagnets in the pre-stress (↓) and post-stress (↓) states are identical, showing no magnetization flipping. **c,** Nanomagnets having a lower shape anisotropy energy (nominal dimension ~200×175×12 nm) than that of stress anisotropy at ~80 MPa experience magnetization switching to ~90º when stress is applied. Upon removal of stress, only a few nanomagnets (due to factors such as lithographic variances, stress distribution and 50% probability of rotating "up" or "down" from the hard axis) flip their magnetization from (↓) to (↑). The yellow arrows identify nanomagnets that experience a flip in magnetization due to the strain-induced switching.

**Figure 2: Clocking of dipole-coupled single-domain magnetostrictive nanomagnets on piezoelectric substrate implementing a Boolean NOT logic gate, and corresponding MFM phase images of elliptical Co nanomagnets on PMN-PT substrate in pre- and post-stress states. a,** A dipole-coupled nanomagnet pair (**L**, **R**) is "initialized" in the 'down' direction by a magnetic field: (*i*) no stress applied, $\sigma = 0$, (*ii*) when stressed with tensile stress, the magnetization of **L** barely rotates owing to its high shape anisotropy while that of **R** rotates by ~90°, to the right due to dipole coupling with **L.** Again, we show the stress producing a coherent 90 degree rotation for the sake of illustration and ease of understanding. The bottom panel shows the potential energy (P. E.) profile as a function of the angle θ subtended by the magnetization of the right magnet with the vertical axis pointing down. The "ball" shows the state of the system under the three different conditions, and (*iii*) when the stress is withdrawn to make $\sigma = 0$, the magnetization of **R** rotates and settles to the 'up' direction. **b,** Dipole-coupled nanomagnet pair (**L** ~250×150×12 nm, **R**~200×175×12 nm) having a separation of ~300 nm in the "initialized", pre-stress state, pointing 'down' (↓↓). Upon stress application (~80 MPa), **R** with the lower shape anisotropy experiences magnetization rotation to ~90°, while **L** experiences no switching

due to its high shape anisotropy. Finally, when stress is removed, the magnetization of **R** rotates and settles to the desired (↑) state, under the dipole influence of **L**, resulting in the final state of the pair as (↓↑).

**Figure 3: A "binary wire" and unidirectional bit information propagation by clocking with strain in an array of three dipole-coupled nanomagnets. a,** An array of three "initialized" dipole-coupled nanomagnets (**L**, **C**, **R**): (*i*) $\sigma = 0$, (*ii*) when $\sigma = \sigma_{max}$, the stress anisotropy overcomes the shape anisotropy energies of **C** and **R** and rotates their magnetizations to ~90°, (we show the stress producing a coherent 90 degree rotation for the sake of illustration) (*iii*) when the stress reduces to an intermediate value, $\sigma_{int} < \sigma_{max}$, the shape anisotropy energy of **C** exceeds that of its stress anisotropy, causing its magnetization to flip and settle to the 'up' orientation as dictated by its dipole interaction with **L**; $\sigma_{int}$ is still high enough to keep the magnetization of **R** at ~90°, (*iv*) Finally, when the stress is removed, the magnetization of **R** rotates and settles to the 'down' direction based on its dipole interaction with **C**. **b,** Three-nanomagnet dipole-coupled pairs (**L**~250×150×12 nm, **C**~200×175×12 nm, **R**~200×185×12 nm). Yellow arrows indicate dipole-coupled nanomagnet pairs that experience magnetization flipping from the "initialized" (↓↓↓) state to the final, settled state (↓↑↓).

# FIGURES

## FIGURE 1

**a**

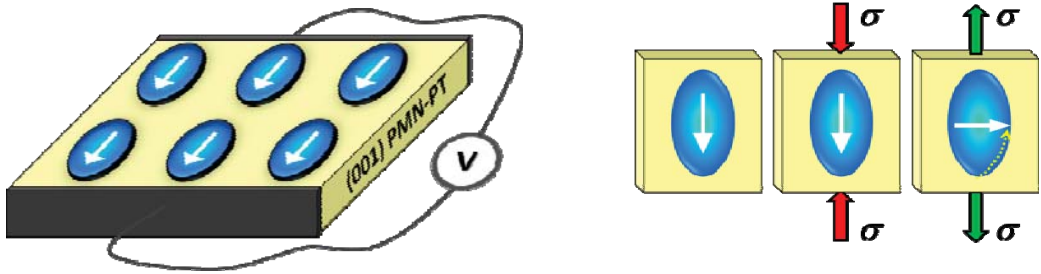

**b**

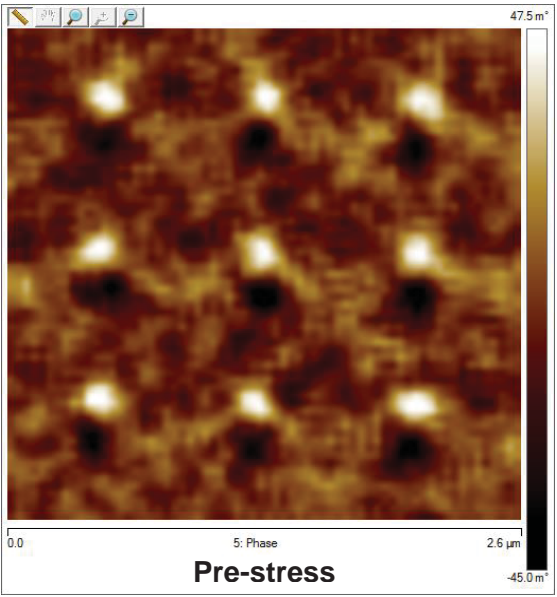
Pre-stress

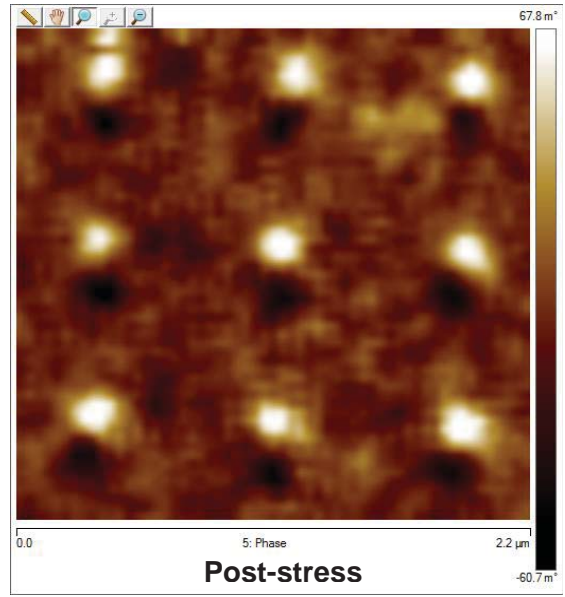
Post-stress

**c**

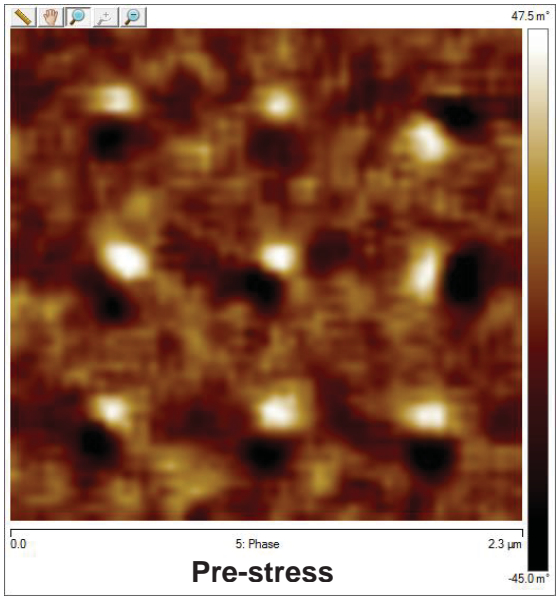 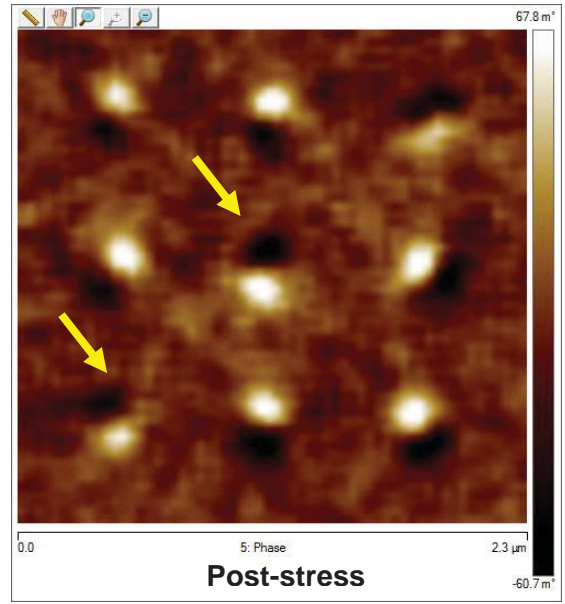

Pre-stress

Post-stress

**FIGURE 2**

a 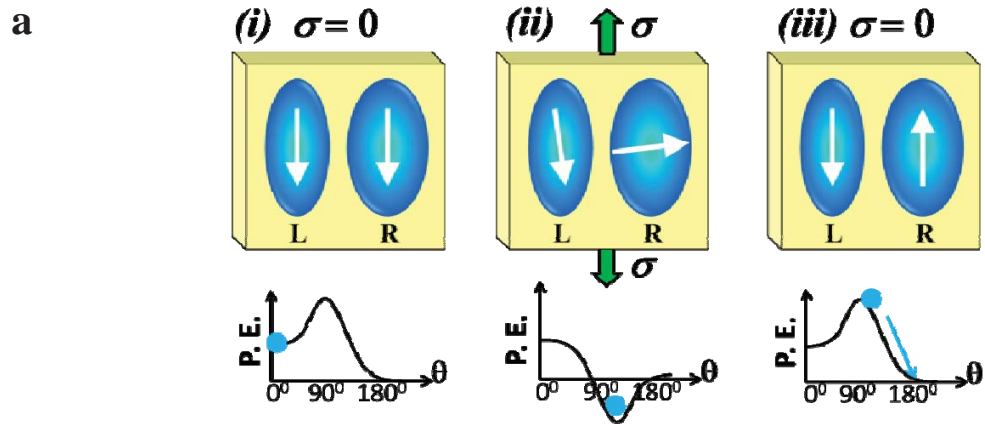

b 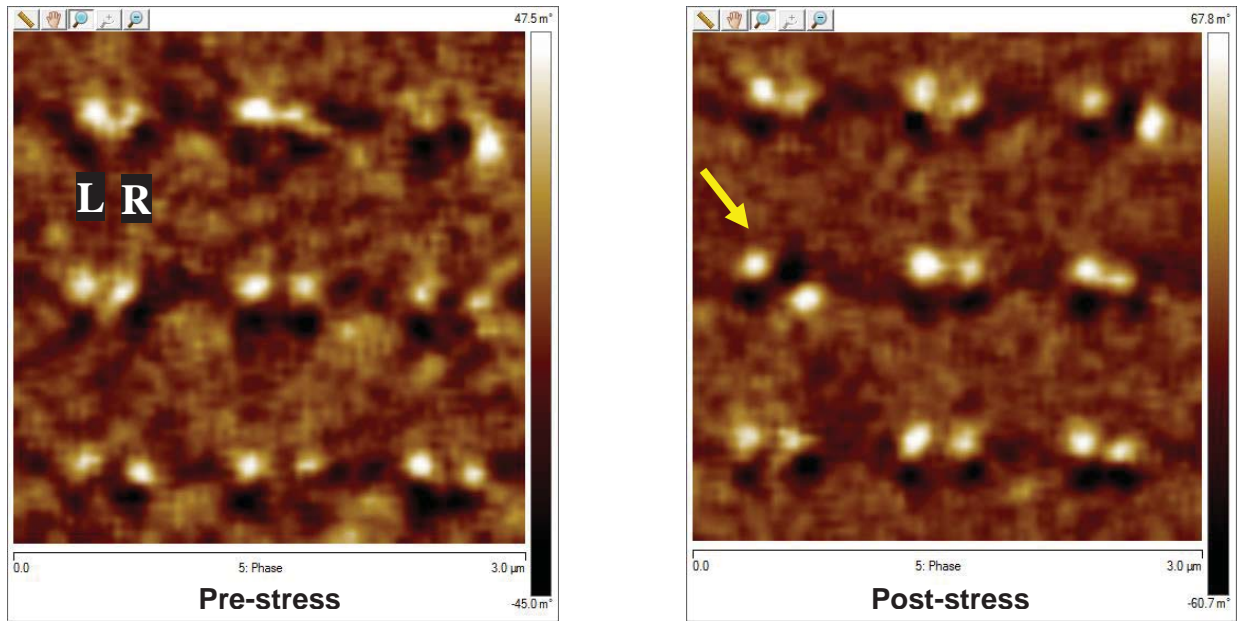

Pre-stress    Post-stress

# FIGURE 3

a 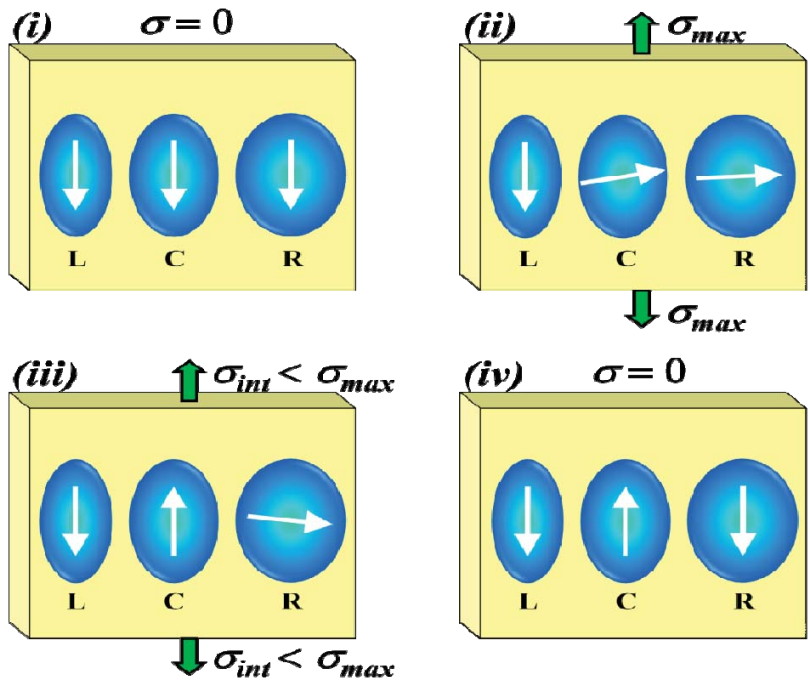

b 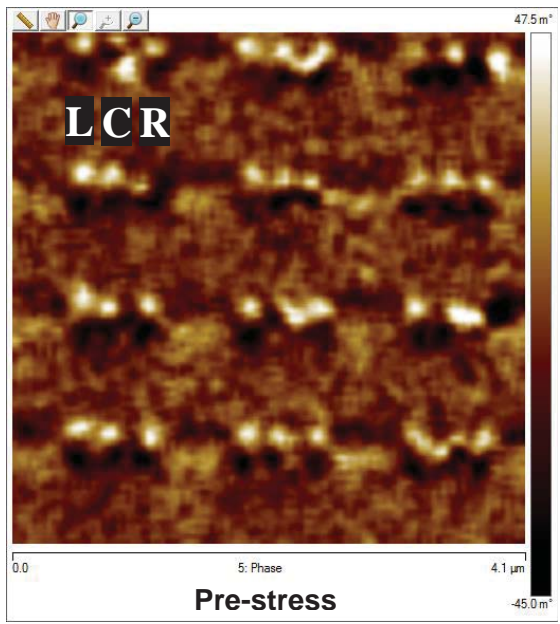 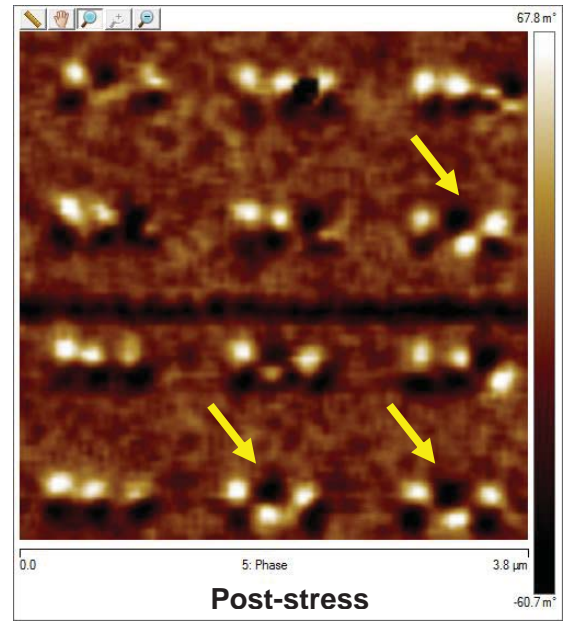

# Experimental Clocking of Nanomagnets with Strain for Ultra Low Power Boolean Logic

Noel D'Souza[a], Mohammad Salehi Fashami[a], Supriyo Bandyopadhyay[b] and Jayasimha Atulasimha[a,*]

[a]Dept. of Mechanical and Nuclear Engineering

[b]Dept. of Electrical and Computer Engineering,

Virginia Commonwealth University, Richmond, VA 23284, USA.

*Email: jatulasimha@vcu.edu

In the main paper, we demonstrated strain-clocked switching of magnetostrictive Co nanomagnets fabricated on a bulk PMN-PT piezoelectric substrate. Owing to the shape anisotropy, the elliptical Co nanomagnets have two stable states for magnetization orientation – 'up' (↑) and 'down' (↓) – along the major axis. Magnetization rotation is accomplished via the Villari effect, or the inverse magnetostrictive effect, in which a strain/stress induces a magnetization rotation in the Co nanomagnets. This strain is produced when a voltage is applied between two electrodes delineated on the PMN-PT substrate. The substrate deforms, generating a strain that is transferred to the magnetostrictive Co layer, which is in elastic contact with the substrate. In this supplement, we present the characterization of the strain developed in the PMN-PT substrate as a function of the applied voltage and the calculation of stress and shape anisotropies in nanomagnets of various nominal dimensions, as well as the sensitivity of these anisotropies to variations in the nanomagnet dimensions. We further present additional experimental data on switching behaviour for the following cases: (i) Isolated magnet, (ii) Dipole-coupled pair, and (iii) Dipole-coupled chain of nanomagnets. Finally, we provide scaling and energy calculations to demonstrate the potential of this paradigm in achieving ultra-low power Boolean computing.

**Supplementary Section A: Experimental Setup and Anisotropy Energy Calculations**

*A.1 Experimental characterization of strain generated in the PMN-PT substrate*

The piezoelectric substrate used in our experiments was a polished (001)-oriented (1-x)[Pb(Mg$_{1/3}$Nb$_{2/3}$)O$_3$]–x[PbTiO$_3$] (PMN-PT) substrate (where x = 0.3) of dimensions 5×5×0.5 mm$^3$ supplied by Atom Optics Co Ltd. In order to measure the strain response of the PMN-PT substrate, we attach a general purpose 120 Ω Constantan linear foil strain gauge (EA-06-062ED-120) Vishay Precision Group, Micro-Measurements) to the top surface of the PMN-PT substrate and measure the strain using a P3 Strain Recorder and Indicator (Vishay Precision Group). Electrodes are attached to the edges of the substrate using silver paste and a voltage is applied along the length of the substrate using a Xantrex XFR20-60 DC power supply in conjunction with a Trek 10/10B high voltage amplifier. Poling is performed in a castor oil bath to prevent arcing at high voltages. An electric field of 800 kV/m is applied along the length of the substrate at a rate of ~1 kV/min. The strain response of the PMN-PT is then measured using the P3 strain recorder, as shown in the strain-voltage curves of Fig. S1. Following PMN-PT poling along the length of the substrate (the direction of P in the inset illustrates the direction of polarization), the strain response is determined for various voltages. It can be seen that for a voltage of 1.5 kV ($E = 300$ kV/m), a strain of ~300 ppm is observed, while at V = 2 kV ($E = 400$ kV/m), a strain of ~400 ppm can be generated. For our numerical calculations, we use the following material constants for a) Co: Young's modulus, $Y = 209$ GPa[1], saturation magnetization, $M_s = 14.22 \times 10^5$ A/m [2], magnetostrictive constant, $(\frac{3}{2}\lambda_s) = -5 \times 10^{-5}$ [2,3]; and b) PMN-PT: Young's modulus, $Y = 105$ GPa, Curie temperature, $T_c = 150$ °C (36). The $d_{33}$ value of (001) PMN-PT experimentally measured in our experiments (~1000 pm/V) is in agreement with other experimentally derived $d_{33}$ values[4–7].

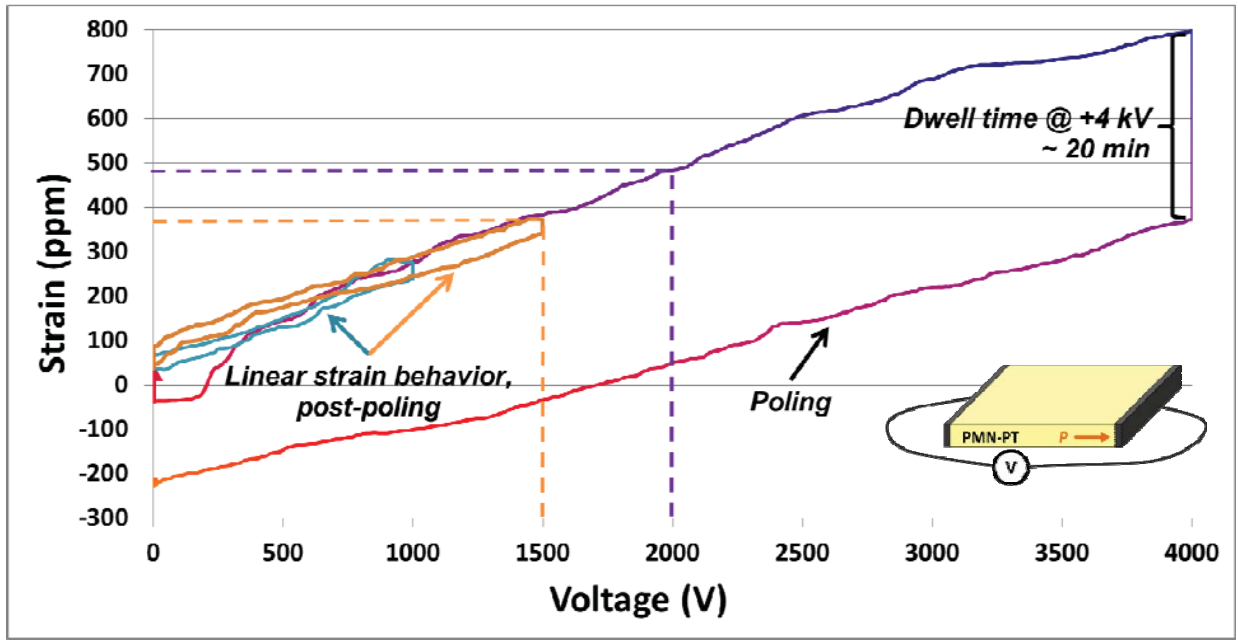

**Fig. S1: Strain response curves for bulk (001) PMN-PT substrate of dimensions 5×5×0.5 mm³.** Poling of the substrate is performed in a castor oil bath with an electric field of 800 kV/m (V = 4 kV). Measurement of the strain response of the poled substrate is then carried out for various fields. A linear strain response can be observed, with a strain of ~300 ppm generated for V = 1.5 kV and ~400 ppm for V = 2 kV.

Thus, if a strain of ~400 ppm is transferred to the Co layer, it corresponds to a stress $\sigma = Y \times$ strain ~80 MPa developed in it. Note that Co nanomagnets are fabricated, as described in the Methods Summary section at the end of this Supplement, on another similarly-poled PMN-PT substrate of the same dimensions (and not on the substrate used in the strain measurements).

The primary considerations in choosing the nominal dimensions of the nanomagnets are: a) in an array of three nanomagnets, the shape anisotropy of the magnets should decrease progressively along the array in order to allow unidirectional propagation of a logic bit, b) the dimension of the "input" magnet (highest shape anisotropy) should be such that its shape anisotropy energy is much greater than the stress anisotropy energy generated by the maximum applied stress of ~80 MPa, so its magnetization will *not* rotate (or rotate very slightly) upon stress application, c) the second magnet (intermediate shape anisotropy) in the array should have dimensions such that a stress of ~80 MPa will be able to produce a ~90° rotation, but a lower stress of ~60 MPa will not. Therefore, when stressed with intermediate stress, its magnetization should rotate to a direction (↑ or ↓) as dictated by its dipole interaction with the neighbouring magnet(s), c) the third magnet (lowest shape anisotropy) should have a lower shape anisotropy than the second magnet so that its magnetization can be rotated by ~90° with a stress of either ~80 MPa or ~60 MPa. Therefore, only when stress is reduced below 60 MPa, should the magnetization of the third magnet (with the weakest shape anisotropy) be able to rotate under the influence of the second magnet, and finally, d) we must account for a ~5% variation in the nanomagnet dimensions, with particular consideration for the second and third nanomagnets, which have smaller tolerances for variations because of the clocking scheme that requires nanomagnets of decreasing shape anisotropy.

Our choice of Co as the magnet material narrows our tolerances further. The higher saturation magnetization of Co ($M_s$ = 14.22 × $10^5$ A/m [2]), compared to that of other common magnetostrictive materials such as Ni ($M_s$ = 4.84 × $10^5$ A/m [2]), enables MFM imaging with better contrast and lower susceptibility to tip-induced magnetization reorientation, but also results in higher shape anisotropy energies for a given set of dimensions. Therefore, the second and third nanomagnets must have shapes that are almost "circular" (low ellipse eccentricity) in order for the stress anisotropy to be able to overcome the shape anisotropy. Consequently, it is extremely important to find a "sweet spot" where the shape anisotropy is sufficiently high to allow good MFM imaging (with low moment MFM tips) but is low enough that the generated stress anisotropy can overcome it and rotate the magnetization. With these objectives in mind, the nominal dimensions of the "input" nanomagnet having the strongest shape anisotropy are chosen to be $(250 \times 150 \times 12)$ nm$^3$, while the second and third nanomagnets are designed with nominal dimensions of $(200 \times 175 \times 12)$ nm$^3$ and $(200 \times 185 \times 12)$ nm$^3$, respectively.

Lithographic and dosage variations make the lateral dimensions of a nanomagnet differ from the nominal values. Deposition rate variation during evaporation of the metals (magnets) makes the thickness random. Another source of variability that is seldom appreciated is oxidation of the Co layer due to repeated handling under atmospheric conditions that reduces the effective dimensions of the nanomagnet (lateral and thickness). In the case of the nanomagnet of nominal dimensions $(250 \times 150 \times 12)$ nm$^3$, a ~5% variation in dimensions (lateral and thickness) will result in lower and upper bound dimensions of $(237 \times 157 \times 11)$ nm$^3$ and $(263 \times 142 \times 13)$ nm$^3$, respectively. Similarly, the lower and upper bounds of the second nanomagnet's dimensions are $(190 \times 183 \times 11)$ nm$^3$ and $(210 \times 167 \times 13)$ nm$^3$, respectively. Finally, the same bounds for the third nanomagnet are $(190 \times 194 \times 11)$ nm$^3$ and $(210 \times 176 \times 13)$ nm$^3$, respectively. It can be seen that for the nanomagnet with weakest shape anisotropy (third), a 5% variation in dimensions results in a 'lower bound' nanomagnet with the easy (long) axis along the horizontal, rather than the vertical, axis!

*A.2 Estimation of the stress anisotropy energy in the Co nanomagnet dots*

Next, we calculate the anisotropy energies of the Co nanomagnets having nominal dimensions of $(250 \times 150 \times 12)$ nm$^3$, $(200 \times 175 \times 12)$ nm$^3$ and $(200 \times 185 \times 12)$ nm$^3$.

The stress anisotropy energy of a nanomagnet can be expressed as[8]:

$$E_{stress-anisotropy} = \left(-\frac{3}{2}\lambda_s\right)\sigma\Omega, \tag{1}$$

where ($\frac{3}{2}\lambda_s$) is the saturation magnetostriction of Co, $\sigma$ is the stress applied to the nanomagnet and $\Omega$ is its volume. A tensile stress is taken to be positive while a compressive stress is negative. Therefore, the stress anisotropy energies of the Co nanomagnets having nominal dimensions as stated above are 8.8 eV, 8.2 eV and 8.7 eV, respectively, for a stress of ~80 MPa in the Co layer.

The stress anisotropy energies associated with the Co nanomagnets of nominal dimensions, as well as with a 5% variation, are displayed in Table 1-1.

**Table 1-1: Stress anisotropy energy of Co nanomagnets**

| Nominal Dimensions | Stress Anisotropy Energy (eV) for Nominal dimensions | Stress Anisotropy Energy (eV) w/ ±5% variation in dimensions |
|---|---|---|
| 250×150×12 nm$^3$ (high shape anisotropy) | ~ 8.8 | (8 – 9.5) |
| 200×175×12 nm³ (low shape anisotropy) | ~ 8.2 | (7.5 – 9) |
| 200×185×12 nm³ (lowest shape anisotropy) | ~ 8.7 | (8 – 9.4) |

*A.3 Estimation of the shape anisotropy energy in the Co nanomagnets*

Next, we calculate the shape anisotropy energy of the nanomagnets which is given by[8]:

$$E_{shape-anisotropy} = \left(\frac{\mu_0}{2}\right)[M_s^2 \Omega] N_d \quad (2)$$

where $\mu_0$ is the permeability of free space, $M_s$ is the saturation magnetization of Co and $N_d$ is the demagnetization factor. We consider the Co nanomagnet to be a very flat ellipsoid (*41*) with the diameters of the major and minor axis as *a* and *b*, and with a thickness *c* (for $a \geq b \gg c$). The expressions for $N_d$ along the major (long) axis and minor (short) axis are[9]:

$$N_{d\_xx} = \frac{c}{a}(1-e^2)\frac{K-E}{e^2}, \quad N_{d\_yy} = \frac{c}{a}\frac{E-(1-e^2)K}{e^2(1-e^2)^{\frac{1}{2}}}, \quad (3)$$

where *K* and *E* are complete elliptical integrals[10] with argument $e = (1-b^2/a^2)^{\frac{1}{2}}$

The shape anisotropy energies associated with the Co nanomagnets of nominal dimensions, as well as with a 5% variation, are displayed in Table 1-2. As can be seen, the shape anisotropy energy of the nanomagnet having the lowest shape anisotropy is still high enough that it would not be affected by random thermal noise at room temperature, thereby minimizing static error probability. However, small variations in dimensions can tip the balance in favour of high shape anisotropy (larger than stress anisotropy energy + dipole energy in case of pairs or coupled arrays) so there would not be many nanomagnets that can switch. Of course, in addition to dimensional variations, pinning sites and defects would also affect the effective barrier for switching.

Table 1-2: Shape anisotropy energy of Co nanomagnets

| Nominal Dimensions | Shape Anisotropy Energy (eV) for Nominal dimensions | Shape Anisotropy Energy (eV) w/ ±5% variation in dimensions |
|---|---|---|
| 250×150×12 nm$^3$ (high shape anisotropy) | ~ 105 | (71 – 148) |
| 200×175×12 nm$^3$ (low shape anisotropy) | ~ 26 | (6 – 53) |
| 200×185×12 nm$^3$ (lowest shape anisotropy) | ~ 16 | (4 – 42) |

One could argue that designing the second and third nanomagnets with even lower shape anisotropy would have ensured that the stress anisotropy would rotate a greater number of nanomagnets. However, consider the third nanomagnet with lowest shape anisotropy having nominal dimensions of (200×185×12) nm$^3$. A 5% variation in every dimension could result in a nanomagnet of dimensions ~ (190×194×11) nm$^3$. It is easy to see that such a nanomagnet would have its easy (long) axis along the horizontal, rather than the vertical, axis, and inhibit propagation of information along the nanomagnet array. Therefore, while nanomagnets with nominal dimensions of, say, (200×190×8) nm$^3$ and (200×195×8) nm$^3$ would have shape anisotropy energies of ~4.7 eV and ~2.3 eV, respectively, and stress anisotropy energies of ~6 eV and ~6.1 eV (generated by 80 MPa stress) will be enough to rotate the magnetization, the possibility of finding nanomagnets with incorrect easy axes (along the horizontal instead of vertical) will also be greater. Furthermore, note that the lower the shape anisotropy, the higher the possibility of tip-induced effects from the MFM tip, which may cause magnetization reorientation during scanning.

Considering the complexities described above, one can appreciate the tight fabrication tolerance of this scheme, especially when considering an array of multiple nanomagnets with decreasing shape anisotropies. Failure to satisfy this strict tolerance accounts for the low percentage of nanomagnets that switch correctly, as shown in the MFM results of Section C of this Supplement. We also point out that such strict lithographic tolerances may be daunting for an academic lab, but is par for the course in an industrial foundry.

## Supplementary Section B: Nanomagnet and Substrate Characterization

a. Nanomagnet characterization

Fig. S2 displays several SEM micrographs to show the quality of the fabricated Co nanomagnets on a PMN-PT substrate. In addition, AFM topography images (Fig. S3) illustrate the quality of the nanomagnets (roughness, thickness and lateral dimensions).

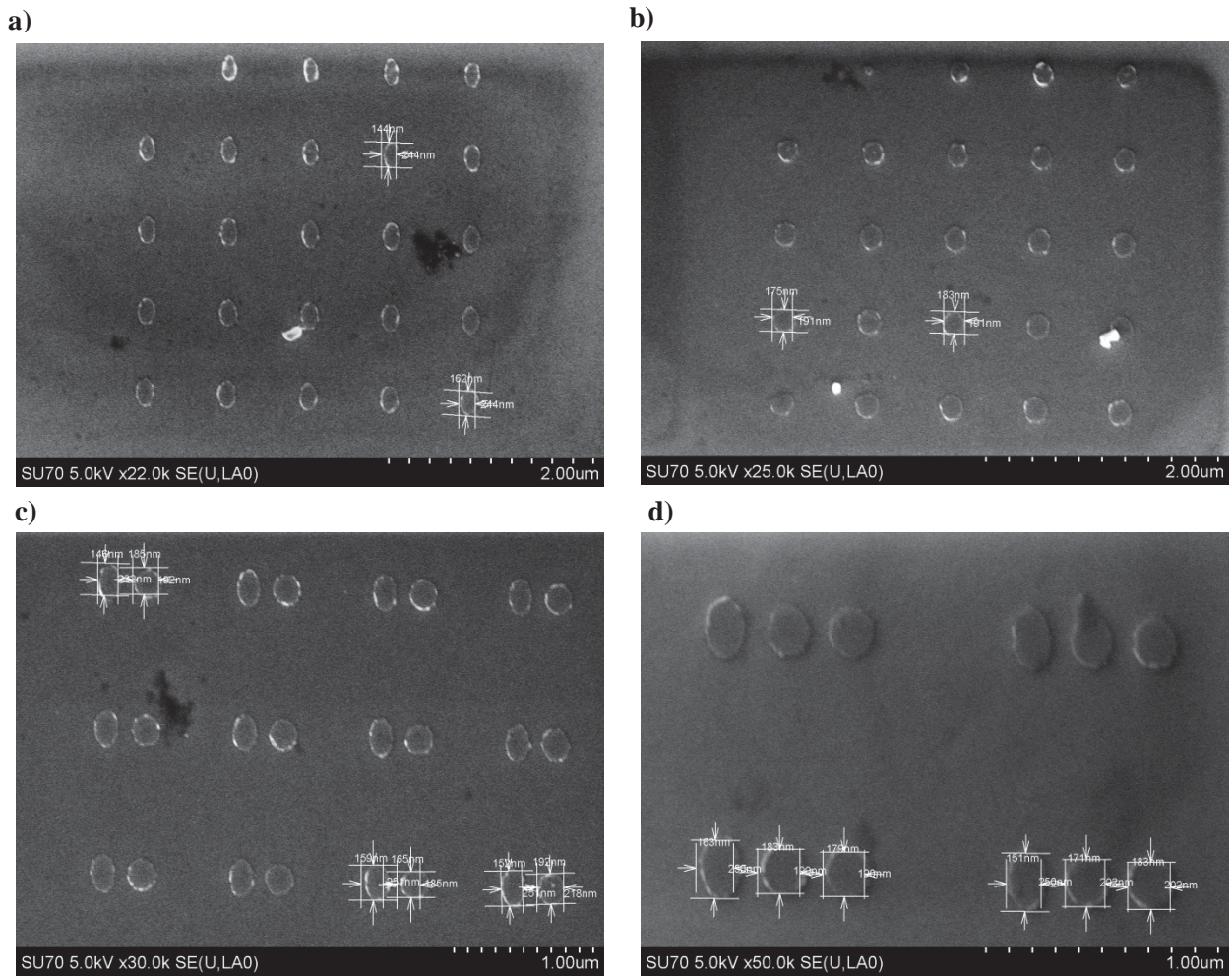

**Fig. S2. Co nanomagnets fabricated on PMN-PT substrate.** The following scenarios are considered, having the corresponding lateral dimensions: a) Isolated, (250 nm x 150 nm), b) Isolated, (200 nm x 185 nm), c) Dipole-coupled with inter-magnet spacing of 315 nm, (250 nm x 150 nm, 200 nm x 185 nm), d) Array with inter-magnet spacing of 315 nm, (250 nm x 150 nm, 200 nm x 175 nm x 185 nm). Thickness = 16.5 nm (5 nm Ti + 11.5 nm Co).

In Fig. S3, the height of one particular nanomagnet is measured to be ~ 15.3 nm. The nominal height of the fabricated nanomagnets is 16.5 nm (5 nm Ti + 11.5 nm Co).

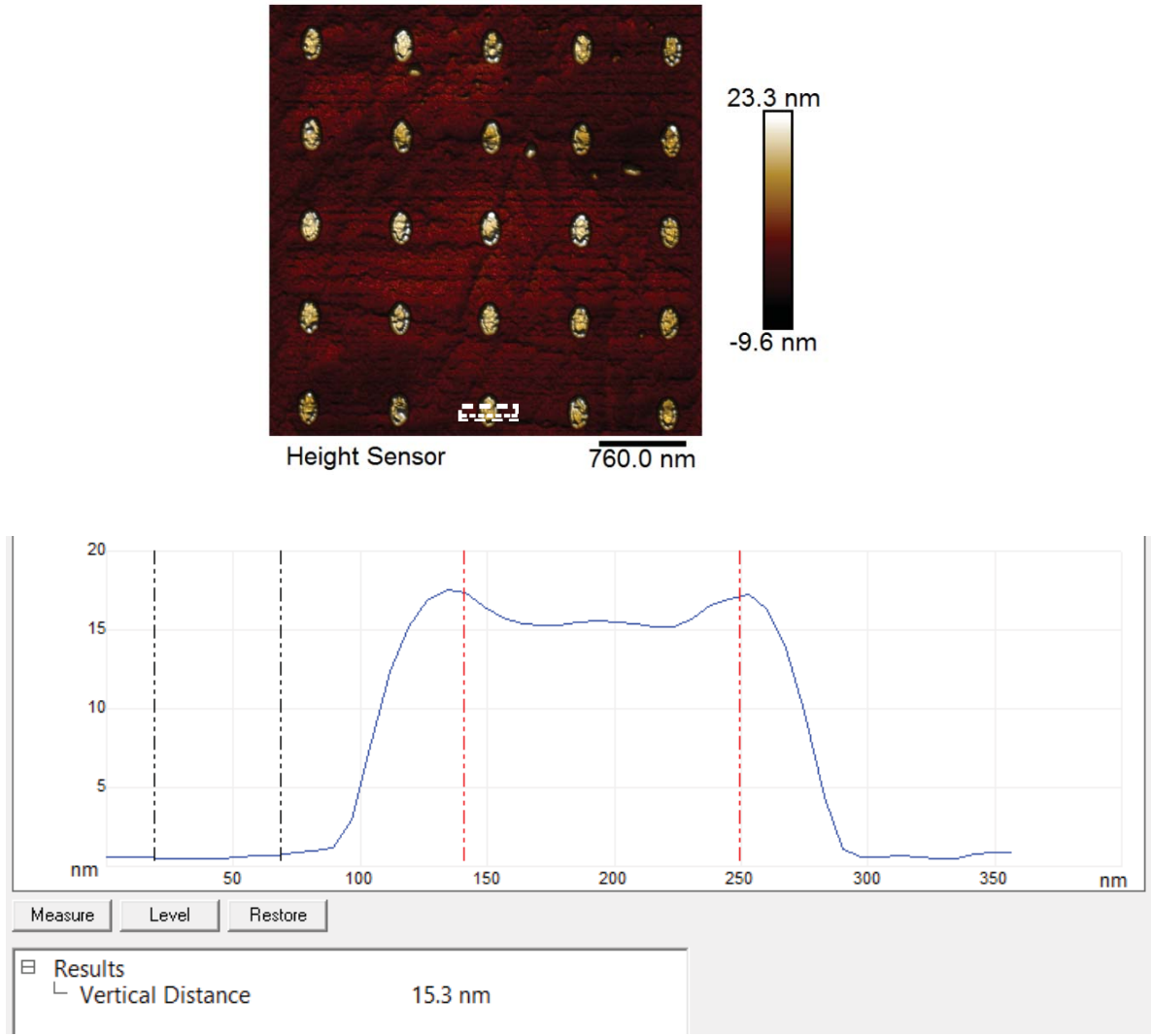

**Fig. S3. Height measurement of Co nanomagnet on PMN-PT substrate. (Note: Total thickness of magnet = 5 nm Ti + 11.5 nm Co).**

b. PMN-PT surface roughness characterization

The surface roughness of the substrate is ~1.7 – 2 nm. As shown in Fig. S4, a 15 μm x 15 μm region of the PMN-PT substrate shows $R_q$ (root mean square roughness) and $R_a$ (average roughness) values of the particular scanned region. A smaller 7.2 μm x 7.2 μm section was also studied since this corresponds to the largest area of our nanomagnet arrays (array of 3 magnets). Local variations in the roughness can rise up to ~3 nm. These surface incongruities are taken into account for potential peculiar switching mechanisms in our MFM studies as such occurrences, though very rare, lead to unclear magnetic states and were discarded from our switching results/claims of correct switching.

a)

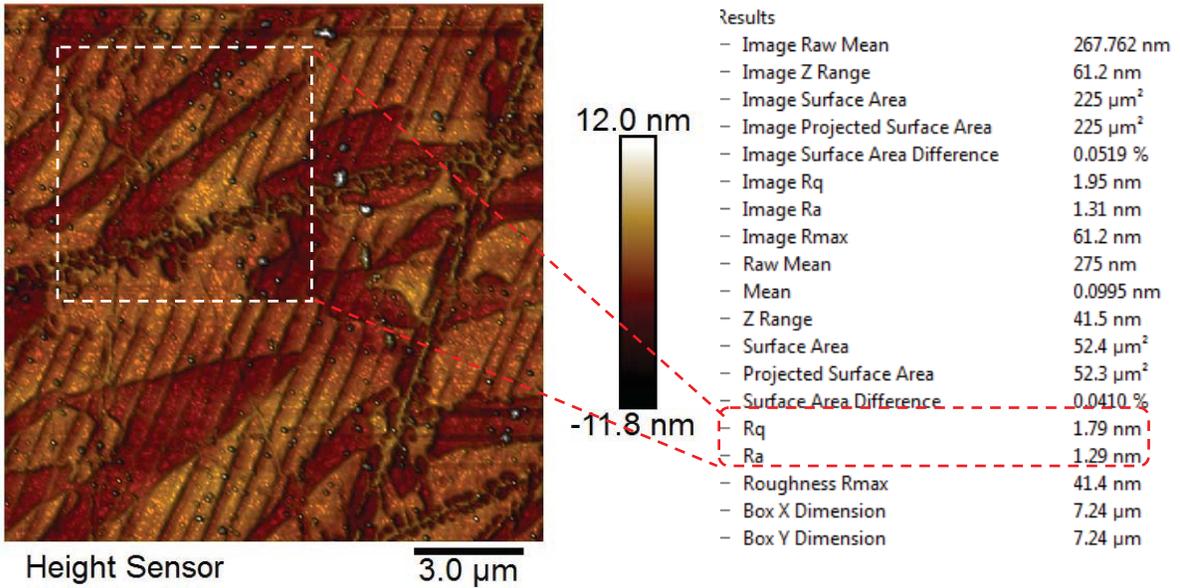

b)

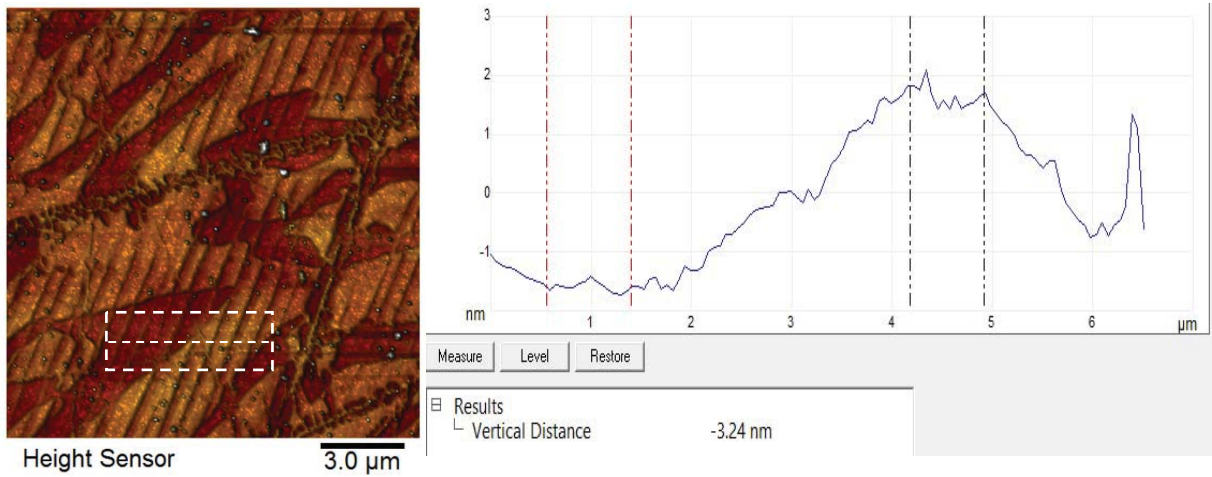

**Fig. S4. a) Surface roughness of PMN-PT substrate in a 15 μm x 15 μm section with the inset highlighting a smaller ~7.2 μm x 7.2 μm sub-section. b) Illustration of local variation in surface roughness.**

c. <u>EDS, XPS analysis of nanomagnets to show oxidation is NOT a critical issue</u>

In order to characterize the films deposited (5 nm Ti + 11.5-12 nm Co), several material characterization techniques were performed.

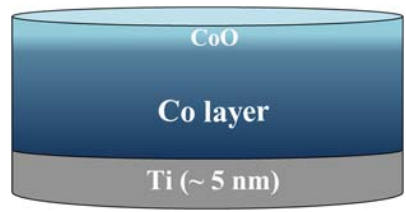

- The Co layer (~11-12 nm) is *not* capped with any material and although this would lead to oxidation of the top 1-2 nanometers of the Co layer, if the experiment is conducted within 2-3 weeks, there will be no significant effect due oxidation on the ferromagnetic behavior of the bulk of the Co nanostructure.

- XPS studies of oxidation conducted on Co (Gan et al., 2003) showed ~1.7 nm of CoO forming after $10^3$ hours. Since our experiments are performed within $10^3$ hours, the resulting anti-ferromagnetic oxidation layer, CoO, would not be thick enough to have a detrimental effect on the magnetization of the metallic ferromagnetic Co layer in the elliptical nanomagnets (Welp et al., 2003). Similar assumptions have also been made by Cui et al. on Ni films (Cui et al., 2013).

- Energy-dispersive x-ray spectroscopy (EDS, using SEM) was performed on the PMN-PT substrate (Fig. S5), as well as on a 11.5 nm Co film subsequently deposited on the PMN-PT substrate (Fig. S6).

Furthermore, the EDS of a 11.5 nm Co thin film deposited on a PMN-PT substrate was specifically performed over 4 days (Figs. S7 and S8, on Si substrate). As can be seen, there is little to no difference in the elemental oxide percentage levels between Day 1 and Day 4.

It should be noted that our surface analysis clearly shows very little oxygen content (Figs. S7 and S8) in thin Co films deposited on Si substrate. On PMN-PT substrate, the apparent high oxygen content is because the penetration depth (~ several microns deep) of the x-ray generating beam focused on to the sample is much greater than the thin film thickness (~12 nm). Therefore, most of the oxygen content in the measured EDS signal is from the PMN-PT substrate and not from the Co layer.

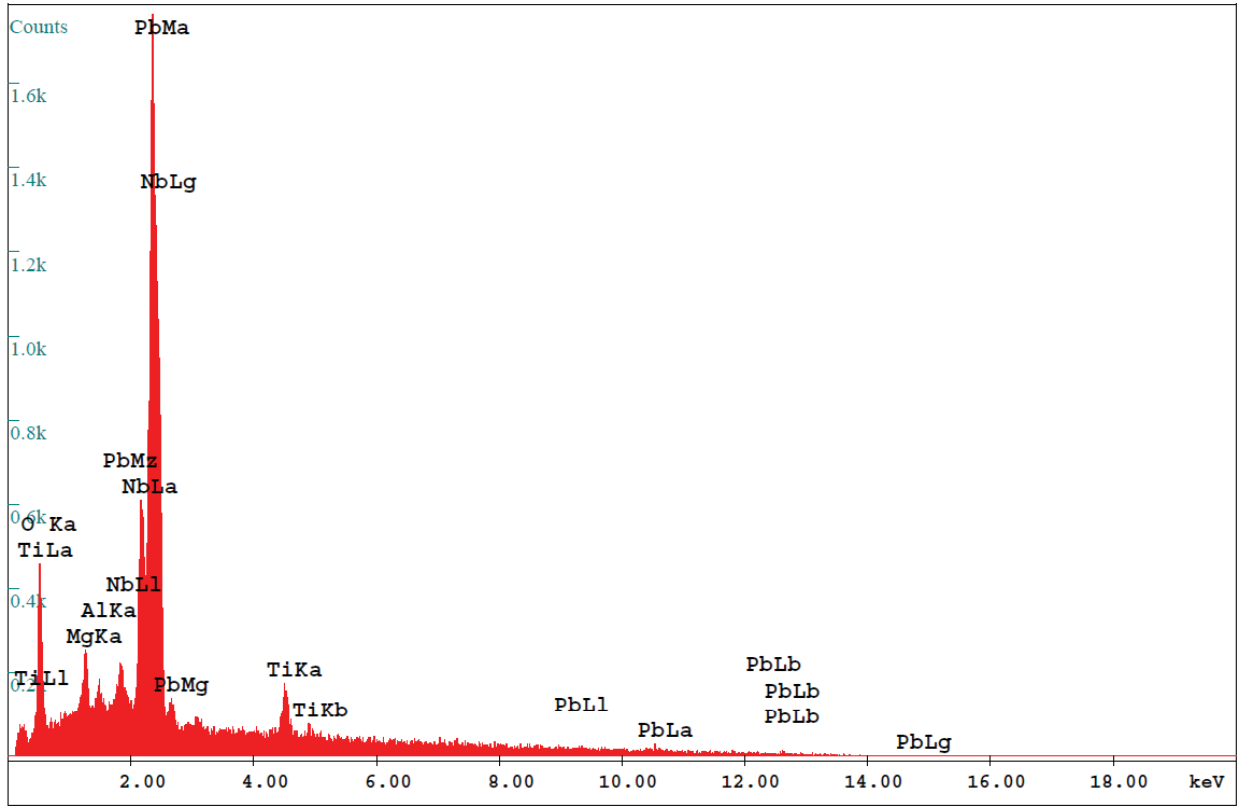

**Fig. S5. EDX results of PMN-PT substrate**

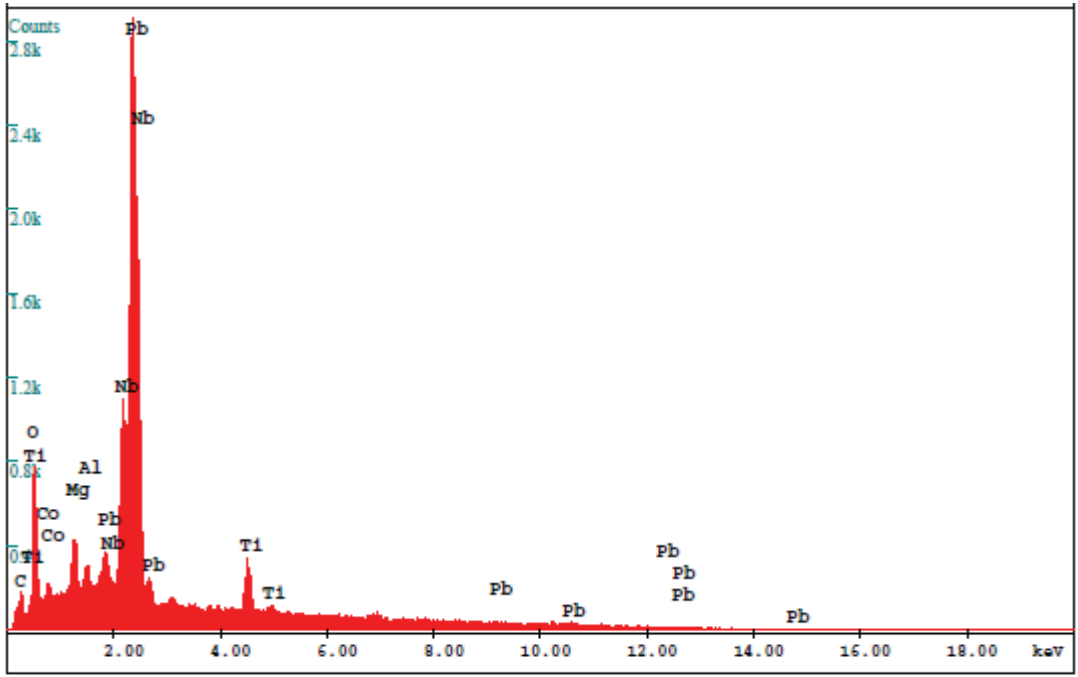

**Fig. S6. EDX results of Co film on PMN-PT substrate**

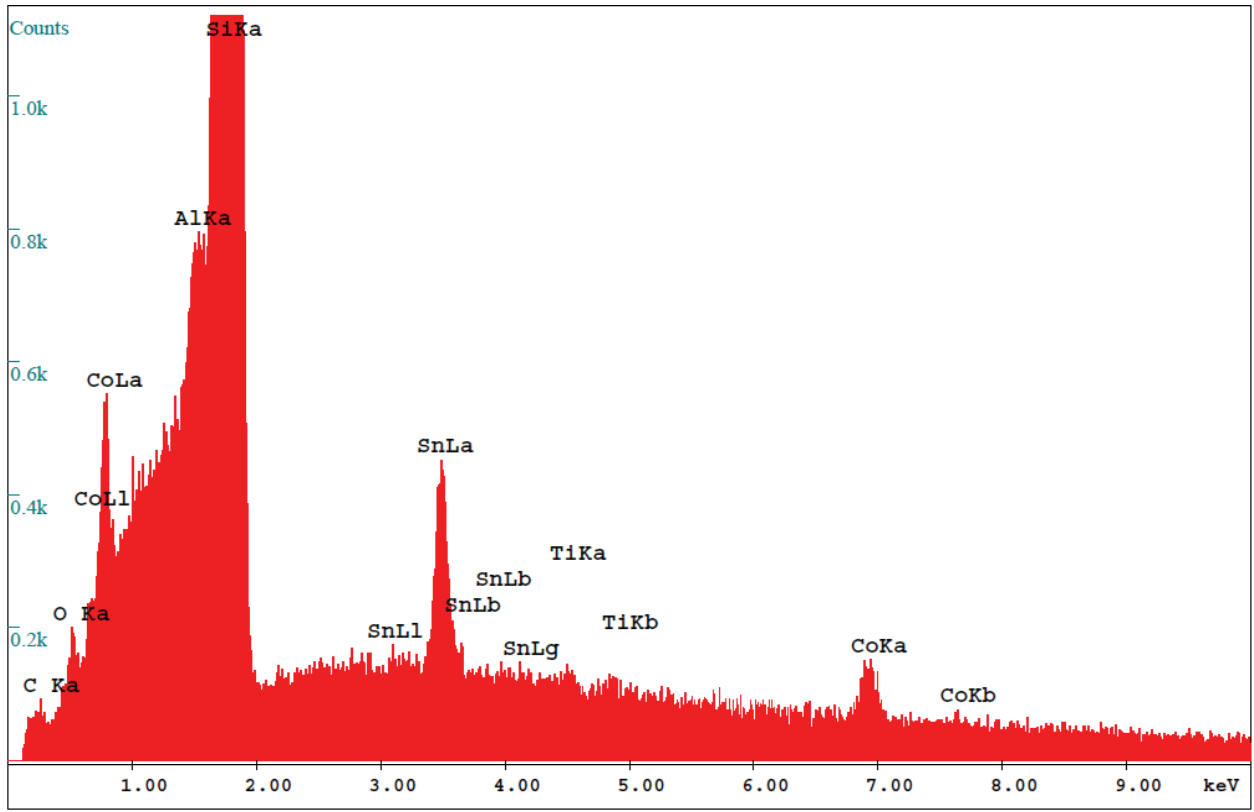

**Fig. S7. EDX results of Co film on Si substrate (Day 1)**

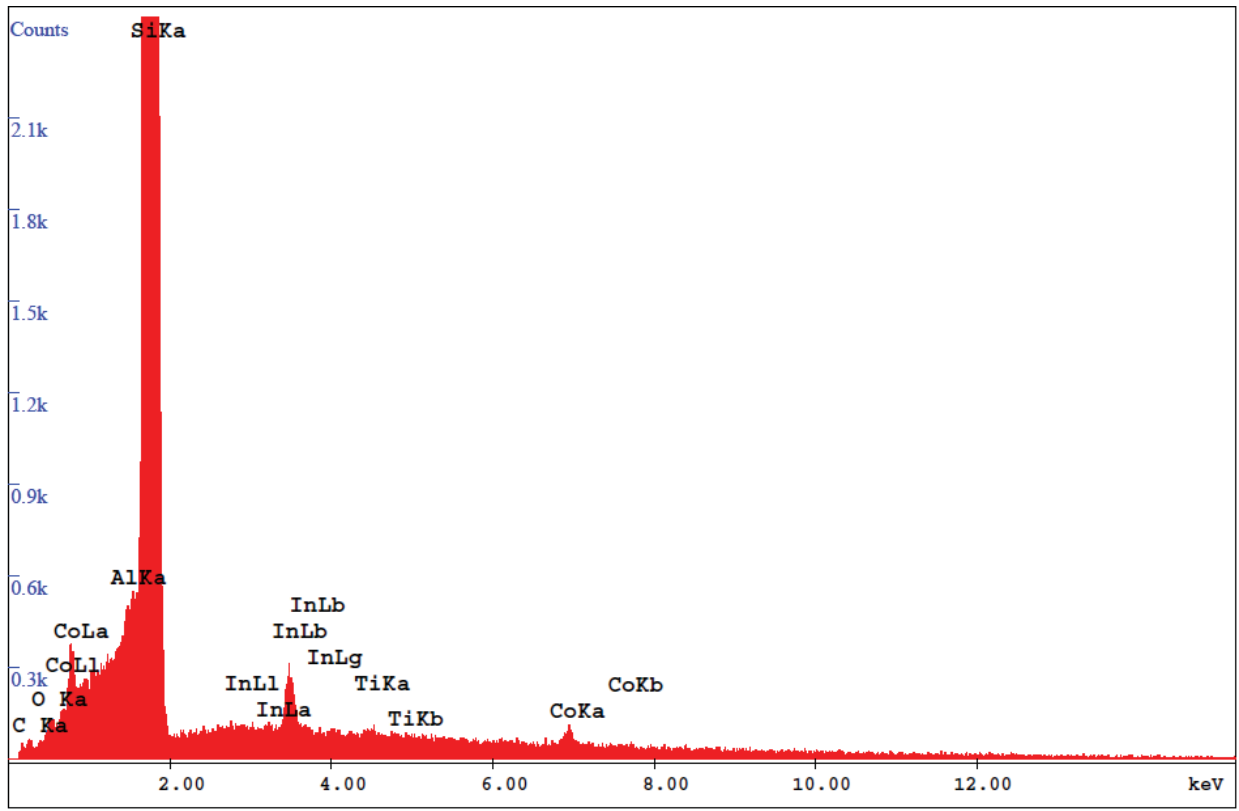

**Fig. S8. EDX results of Co film on Si substrate (Day 4)**

d. Hysteresis (M-H) loops of thin-film Cobalt

In order to characterize the magnetic hysteresis of the cobalt used to fabricate our nanomagnets, M-H plots of a thin, 12 nm Co film were generated using a Quantum Design Versalab™ Vibrating Sample Magnetometer (VSM) at room temperature.

a)
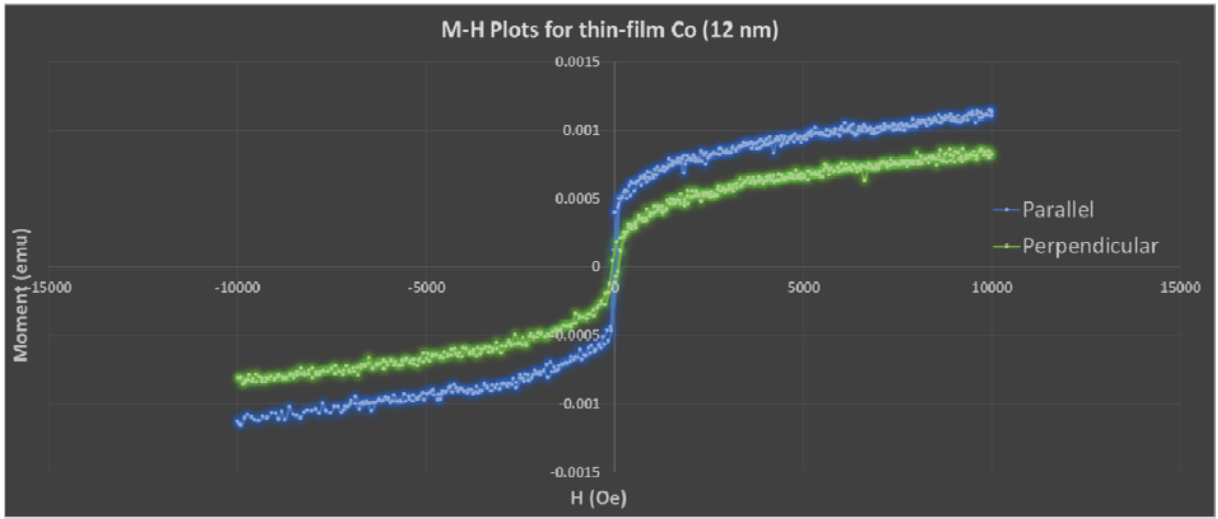

b)
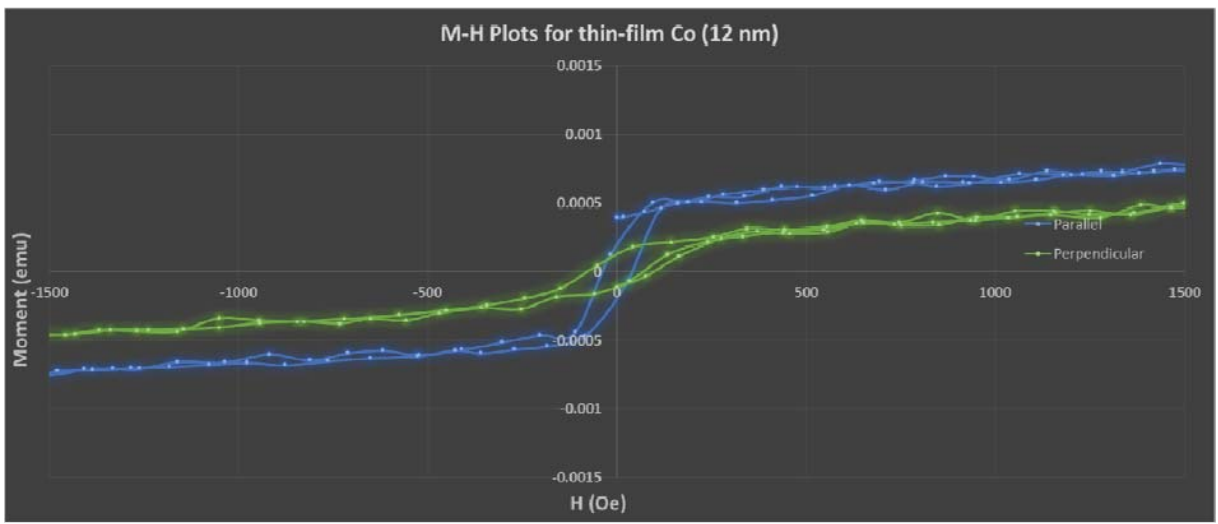

**Fig. S9. Hysteresis plots for 12 nm Co film. a) -10,000 Oe to +10,000 Oe. b) -1,500 Oe to +1,500 Oe**

The effective sample size was ~2 mm x 3 mm. Figure S9a illustrates the M-H loops of the thin-film cobalt when subjected to a magnetic field from -10,000 Oe to +10,000 Oe. Two scenarios are plotted: i) H-field applied along the surface of the substrate (parallel, blue curve), and ii) H-field applied perpendicular to the face of the substrate (perpendicular, green curve). It can be seen that it is easier to magnetize the cobalt parallel to the surface than perpendicular to it. Thus, the "easy" axis of the magnetization is in-plane, or along the surface of the sample. The sample does not have perpendicular magnetic anisotropy (PMA).

**Supplementary Section C: Magnetization Switching Results**

a. Magnetic Force Microscopy (MFM) – Cycle 2 (Cycle 1 performed in main paper)

In the main paper, we present various magnetic force microscopy (MFM) images that illustrate magnetization switching in cases where the stress anisotropy energy of certain nanomagnets is greater than their shape anisotropy energy. Several scenarios are studied: a) nanomagnets with negligible dipole interaction (Case I), b) two dipole-coupled magnets (Case II), and c) array of three dipole-coupled magnets (Case III). These results represent the magnetic states before application of stress (nanomagnets are "initialized" to have their magnetizations point in the 'down' direction (↓) via a strong magnetic field) and after applying one cycle of stress ~80 MPa (Cycle 1). In order to test the repeatability of the magnetization switching demonstrated in the main paper, we perform another MFM study (Cycle 2) on the *same nanomagnet arrays* in which we re-"initialize" the magnetization, in this instance to (↑) with a strong magnetic field of ~200 mT directed along that direction. It is this Cycle-2 data that is presented in this section. After removing the field, we record the magnetic state at zero stress, then apply a strain of ~400 ppm that would produce a stress of ~80 MPa and capture the final magnetization orientation. In the following MFM images, we compare the pre- and post-stress magnetic states of nanomagnets in the three scenarios in Cycle 2.

Another issue that we must confirm – which fortunately did not occur in these experiments – is MFM tip-induced magnetization reorientation in the Co magnets. We perform several consecutive scans of the same nanomagnet array (top-down scan followed by bottom-up scan, and so on). Since no switching occurs owing to scanning, we conclude that the magnetization of the MFM tip is not strong enough to affect the magnetization of the nanomagnets.

Note that the same nanomagnet arrays are investigated in both Cycle 1 and Cycle 2 for all three scenarios. Also, a small amount of nanomagnet sets appear to have contaminants on the surface after Cycle 2, possibly from contaminant accumulation on the MFM tips or from repeated applications of silver paste along the substrate edges. The nanomagnets affected by these contaminants are not considered in our conclusions about magnetization switching.

*C.a.1: Isolated nanomagnets*

In Fig. S10, we show MFM images of isolated nanomagnets with negligible dipole interaction (~800 nm inter-magnet separation). As in Fig. 2a (main paper), we see that the magnetization directions of the highly shape anisotropic nanomagnets do not flip after applying a stress ~80 MPa, as shown in the identical pre-stress (↑) and post-stress (↑) MFM phase images of Fig. S10a. Nanomagnets with lower shape anisotropy (nominally 200×175×12 $nm^3$) are shown in Fig. S10b and we do observe magnetization rotation from (↑) to (↓) (yellow arrows), although these are not the same nanomagnets that switched in Cycle 1 (green arrows). This can be attributed to the fact that the stress induces a magnetization rotation (to the hard axis) in these nanomagnets, but once the stress is removed, there is a 50% probability of the magnetization rotating in either direction since they are under no (or negligible) dipole influence. Thus, a magnet that switched the first time need not switch the second time and vice-versa. Nanomagnets having nominal dimension of 200×185×12 $nm^3$ are shown in Fig. S10c. This figure illustrates magnetization rotation in the nanomagnet identified by the yellow arrow. As can be seen, the magnetization direction of several nanomagnets is not strictly (↑) and tends to be slightly deviated from the vertical direction (also in the case of Fig. S10b). This is due to variations in lithographic fabrication that result in nanomagnets having slight skewed asymmetries in their shape. Therefore, the major (easy) axis may be slightly slanted and not strictly along the vertical axis as desired.

*C.a.2: Dipole-coupled pair*

We also investigated dipole-coupled nanomagnets consisting of a highly shape-anisotropic "input" nanomagnet (~250×150×12 nm; left) that does not rotate significantly under stress and a less shape-anisotropic "output" nanomagnet (~200×175×12 nm; right) whose magnetization does rotate when stressed. It can be seen that two pairs of dipole-coupled nanomagnets (identified with yellow arrows) rotate from the initial (↑↑) state to the final (↑↓) state, indicating a flip in the output magnetization state upon application of stress (Fig. S11a). Interestingly, the nanomagnet pair identified by the green arrow also exhibited magnetization switching in Cycle 1 (in which the rotation was from its pre-stress state of (↓↓) to a post-stress state of (↓↑)). Also, in order to ensure that the MFM tip does not induce magnetization rotation in the nanomagnets, we perform three consecutive scans (top-down, followed by bottom-up scans, and finally another top-down scan) of the same array shown in Fig. S11a. Since all three scans are identical (Fig. S11b), we can conclude that the MFM tip has a negligible effect on switching the magnetization of the nanomagnets.

*C.a.3: Dipole-coupled chain of nanomagnets*

In Fig. S12, we examine an array of three dipole-coupled nanomagnets of decreasing shape anisotropy and having nominal dimensions of 250×150×12 $nm^3$ (left), 200×175×12 $nm^3$ (centre) and 200×185×12 nm³ (right) with an inter-magnet separation of ~300 nm. As before, a global magnetic field (~200 mT) is applied to the nanomagnet arrays to "initialize" the nanomagnets to (↑↑↑). However, lack of precise lithographic control caused some nanomagnet dimensions to differ from the nominal dimensions. As a result, certain nanomagnets may have nearly circular shape with shape anisotropy energies that are lower than the dipole interaction energy due to their neighbours. In these cases, magnetization switching occurs as soon as the initializing magnetic field is removed, and before any stress is applied, because the dipole interaction between neighbours can overcome the small shape anisotropy energy barrier of the nearly-circular magnet and flip its magnetization without the aid of stress (no clocking required). This situation is identified by the red arrows in Fig. S12, which show trios with initial pre-stress states of (↑↓↑) instead of (↑↑↑). The yellow arrow in Fig. S12a identifies a trio in which stress induces a magnetization rotation from its initial state (↑↑↑) to the desired final state (↑↓↑). In another magnet trio (blue arrow), the initial magnetization state is (↑↑↓). However, after applying the stress, the final state of the array is the desired state (↑↓↑). This signifies that when stress was applied, the magnetization of both the central and the right magnets get reoriented to the correct state based on dipole interactions with the "input" magnet on the left having the highest shape anisotropy (thereby, being marginally affected by stress). In another trio, with similar shape anisotropy variation in the nanomagnets (Fig. S12b), we see correct magnetization switching from (↑↑↑) to (↑↓↑) after application of stress (yellow arrow) and from (↑↓↓) to (↑↓↑) (blue arrow). However, we also see instances of seemingly incorrect switching from (↑↑↑) to (↑↑↓) (white arrow). This may be due to several factors such as lithographic variances that result in the central nanomagnet having higher shape anisotropy than desired, stress variation in the substrate, etc. The green arrows identify nanomagnet arrays that switched in Cycle 1, with the dotted white box highlighting the set of arrays investigated in the main paper. It can be seen that neither of the three nanomagnet trios (green arrows) switched in Cycle 2. This can be attributed to the fact that the initial pre-stress state was probably incorrect and not the desired (↑↑↑) state.

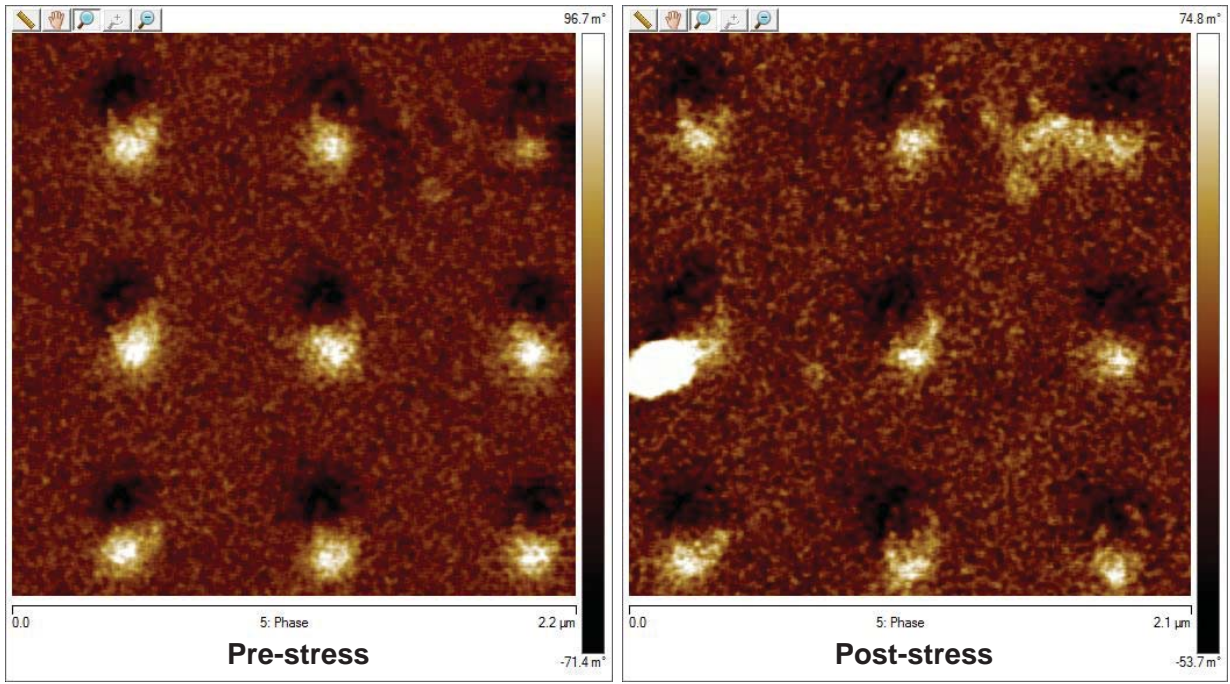

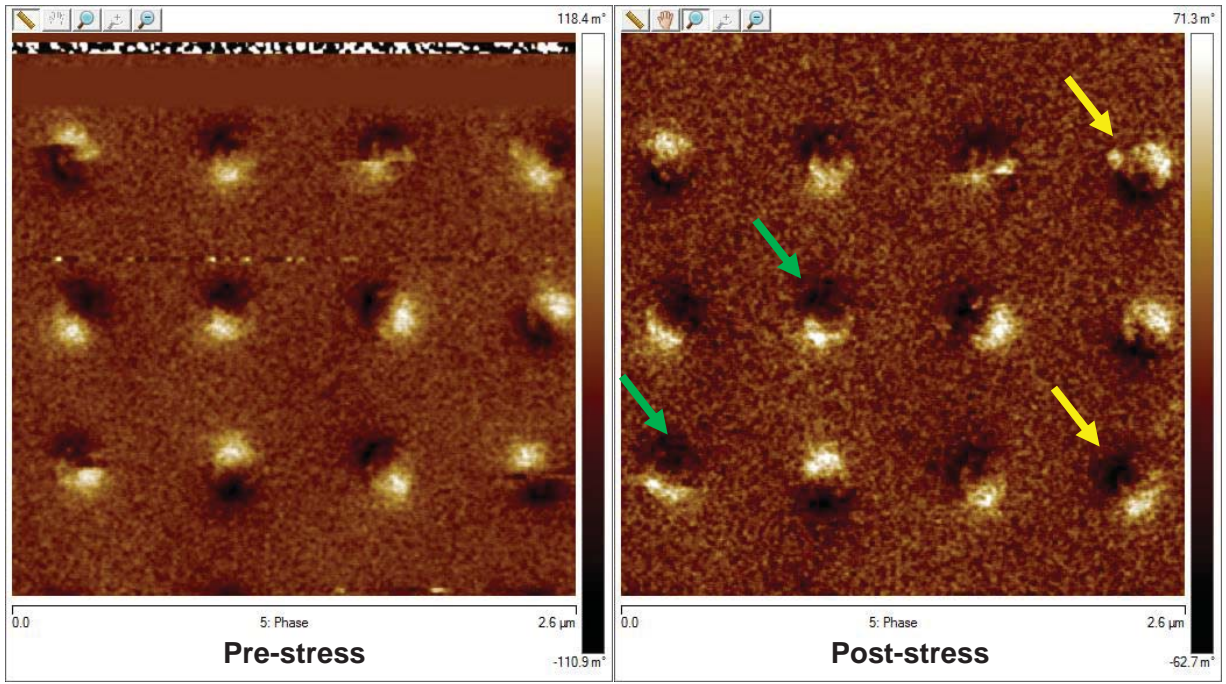

**c)**

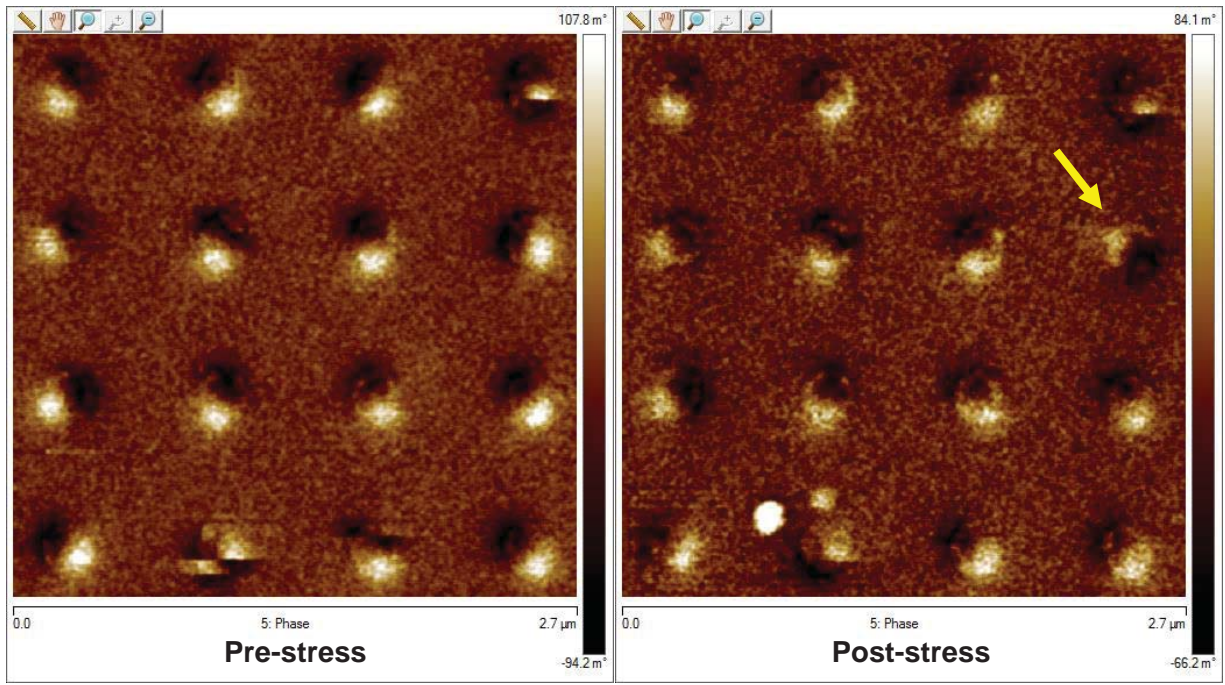

**Fig. S10: MFM phase images of Co nanomagnets on bulk PMN-PT substrate with negligible dipole interaction with neighbours, in pre- and post-stress states.** The nanomagnets are "initialized" to (↑) with a magnetic field of ~200 mT. **a,** Highly shape anisotropic nanomagnets (nominal dimensions ~250×150×12 nm$^3$). Since the shape anisotropy energy is much higher than the stress anisotropy energy (at ~80 MPa), the nanomagnets do not respond to stress and flip. Thus, the post-stress magnetization state of the nanomagnets (↑) is identical to that of the pre-stress state (↑) for all the nanomagnets. **b,** Nanomagnets of lower shape anisotropy (~200×175×12 nm$^3$). When a stress of ~80 MPa is applied, magnetization rotation of ~90° takes place in those nanomagnets in which the stress anisotropy energy exceeds the shape anisotropy energy. When the stress is withdrawn, the magnetizations of these nanomagnets have a 50% probability of flipping from (↑) to (↓), with the yellow arrows highlighting such a scenario. The green arrows point to the nanomagnets that flipped their magnetization in Cycle 1, but not in Cycle 2. **c,** Nanomagnets having the lowest shape anisotropy in our experiments (~200×185×12 nm$^3$). The yellow arrow shows the nanomagnet undergoing magnetization switching.

a)

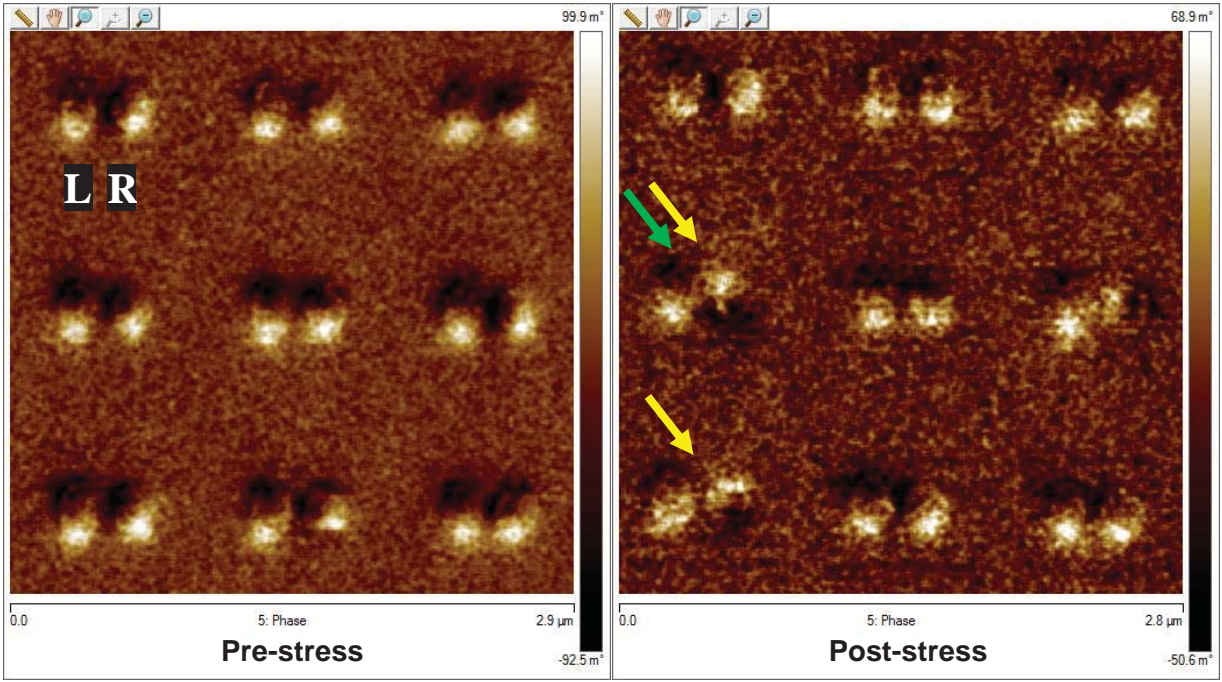

Pre-stress

Post-stress

b)

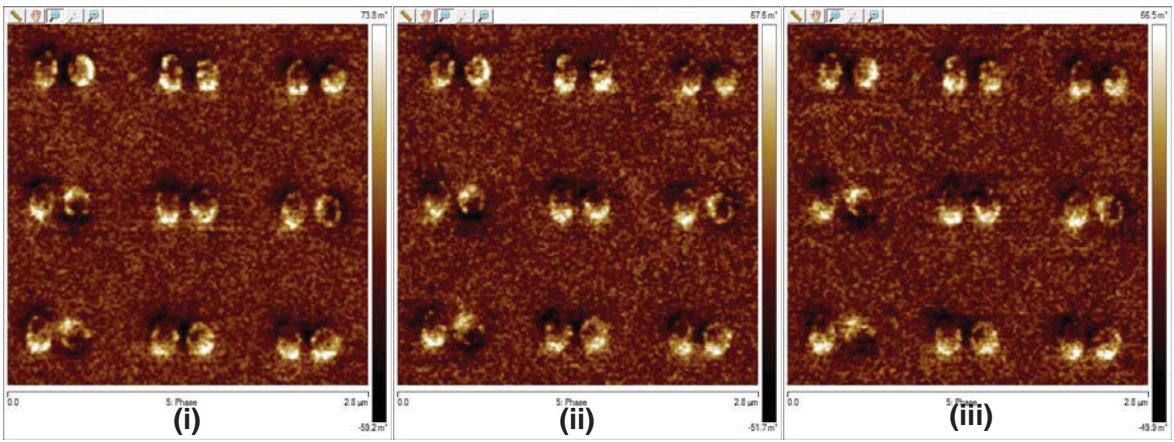

(i) (ii) (iii)

**Fig. S11: MFM phase images of dipole-coupled Co nanomagnets on bulk PMN-PT substrate with dipole interaction between neighbours in pre- and post-stress states. a,** Nanomagnet pairs (**L**~250×150×12 nm$^3$, **R**~200×175×12 nm$^3$) with separation of ~300 nm between their centers. The initial state of the pairs is (↑↑) enforced with a magnetic field. Upon stress application of ~80 MPa, the magnetization of the "output" magnet **R** rotates by ~90° since the stress anisotropy energy is greater than its shape anisotropy energy barrier, while that of "input" **L** undergoes no significant rotation owing to the high shape anisotropy. When the stress is withdrawn, the magnetization of **R** rotates to the (↓) direction as dictated by its dipole interaction with **L**. This scenario is highlighted by the yellow arrows. Other nanomagnet pairs do not undergo this desired switching behavior, possibly due to variations in the fabrication process. The green arrow shows the nanomagnet pair that underwent magnetization switching in Cycle 1 as well [from (↓↓) to (↓↑)]. **b,** Consecutive MFM scans [(i) top-down, (ii) down-top, (iii) top-down] of the nanomagnet array of Fig. S11a. The identical states in all three cases confirm little or no tip-induced magnetization reorientation.

**a)**

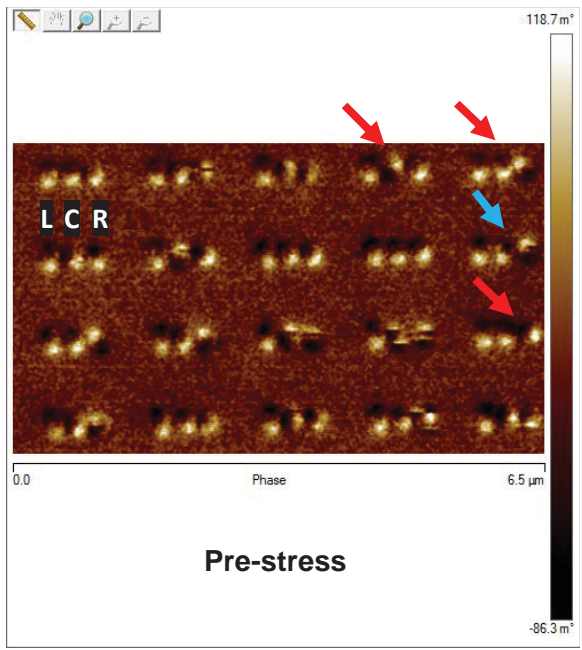 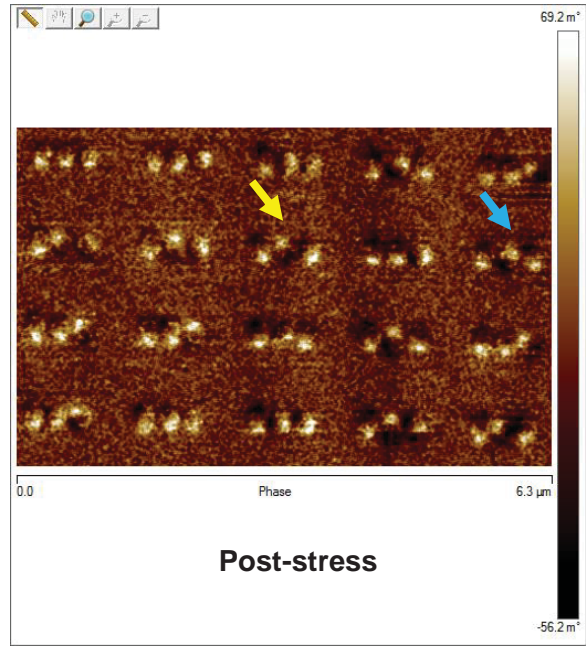

**b)**

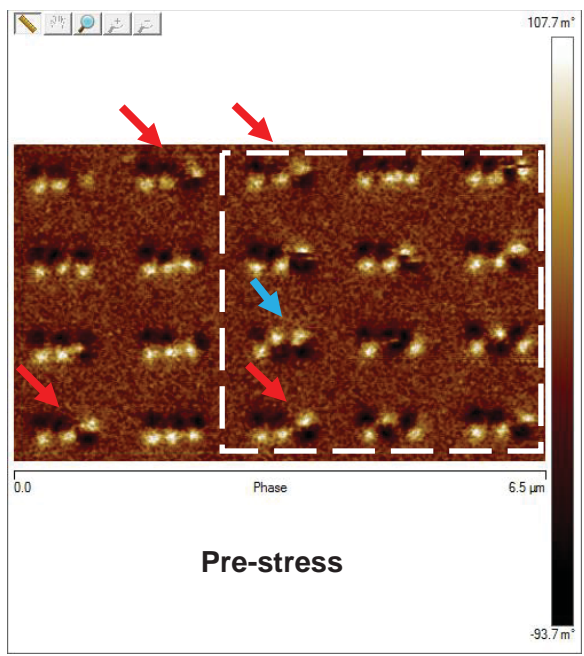 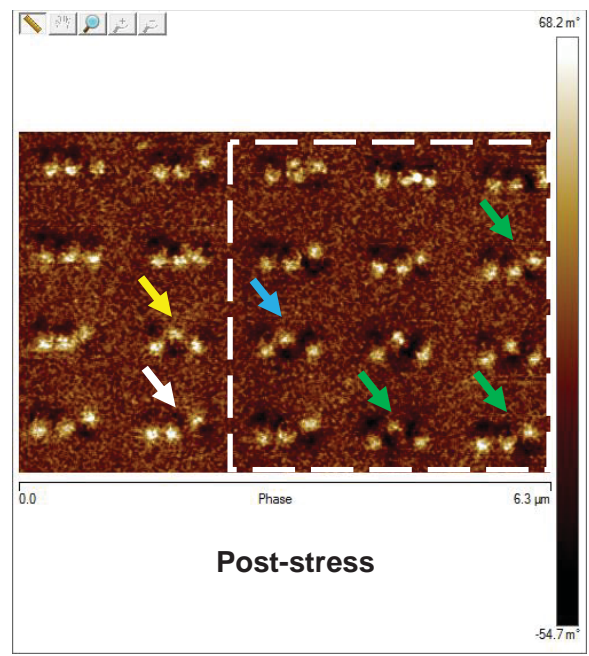

**Fig. S12: Dipole-coupled nanomagnet array consisting of three nanomagnets of decreasing shape anisotropy**. (**a, b),** Nanomagnets (**L**, **C**, **R**) with nominal dimensions ~ 250×150×12 nm$^3$, 200×175×12 nm$^3$, 200×185×12 nm$^3$, respectively. The nanomagnet arrays are "initialized" to (↑↑↑) with a magnetic field. However, certain arrays have incorrect pre-stress initial states (red arrows), possibly due to lack of lithographic control that result in nanomagnets having shape anisotropy energies that are less than the dipole interaction energies they experience. That causes magnetization switching as soon as the initializing magnetic field is removed, and before any stress can be applied. The yellow arrow pinpoints arrays undergoing correct magnetization switching from (↑↑↑) to (↑↓↑). The blue arrow points to an array with incorrect initial states that settle to the desired final state of (↑↓↑) after application of stress ~80 MPa. The white arrow points to another array having a correct initial state but an incorrect final state of (↑↑↓) after applying stress. The green arrows identify nanomagnet arrays that switched in Cycle 1, with the dotted white box highlighting the set of arrays investigated in the main paper.

b. <u>Switching "statistics" on other samples</u>

While the above MFM and magnetization switching studies (Cycle 1 in the main paper, Cycle 2 in the previous section) were performed on the same sample, other such nanomagnets were fabricated on multiple other PMN-PT substrates and investigated.

It must be noted that while all the fabricated samples consist of multiple nanomagnets in multiple arrays, the focus of our studies is on demonstrating magnetization reversal due to strain in <u>a particular nanomagnet</u> (or nanomagnet pairs for dipole-coupled NOT logic and information propagation in three nanomagnet arrays) in a deterministic and repeatable manner.

In a particular array of 9 nanomagnets (as illustrated in our results), in which we expect magnetization rotation due to stress, the 'nominal' dimensions are chosen so that the shape anisotropy is *less than* the stress anisotropy. However, post-lithography and lift-off, there is a slight deviation from the 'nominal' dimensions. Therefore, out of the 9 nanomagnets, there is a fraction in which the shape anisotropy becomes *greater than* the stress anisotropy. We deem this an issue of fabrication-related limitation, rather than a fundamental issue with regard to the physics of the switching behavior. The SEM images of Fig. S13 show examples of the variation in lateral dimensions of the fabricated Co nanomagnets on a PMN-PT substrate. It is clear that many nanomagnets would have huge shape anisotropy due to fabrication imperfection (deviation from nominal dimensions) that prevents them from switching.

Consequently, the 'yield' of observable magnetization switching in these fabricated nanomagnets is not the main focus of our studies. The primary goal is to investigate switching recurrence (through multiple stress cycles) in those nanomagnets that *do* show magnetization rotation. The fact that high error rates and fabrication tolerances affect the yield of switching does not detract from the underlying physics driving this scheme. This is also shown in the observation that switching events (albeit low in number) occur in every sample tested.

The important aspect, therefore, is how often the switching takes place in a particular nanomagnet (or dipole coupled pair) once we have identified the nanomagnet(s) whose shape anisotropy is less than the stress anisotropy and, therefore, are expected to switch consistently.

Multiple cycles of "initialization" (with a magnetic field) and electric field/stress application were scheduled in order to study important aspects of the magnetization switching, such as repeatability, switching statistics, randomness of switching, etc. However, due to the frailty of the PMN-PT substrate, especially when subjected to repeated cycles of high electric fields, there is inevitable crack formation in the substrate which ultimately causes sample failure. As a result, substrate degradation used to occur after 2-3 cycles of stress application (due to the large electric fields applied to the substrate).

One particular method of poling the PMN-PT substrates, involving elevated temperature of the oil bath in which we immerse the PMN-PT during poling (F. Li et al., 2014), minimized the amount of post-poling cracks within the substrate by reducing the large strain variation of ferroelastic domain switching. While this seems to prevent crack formation after poling, it does not prevent eventual substrate degradation after several cycles of electric field application (2-3 stress cycles). This prevents the compilation of an extensive set of switching statistics.

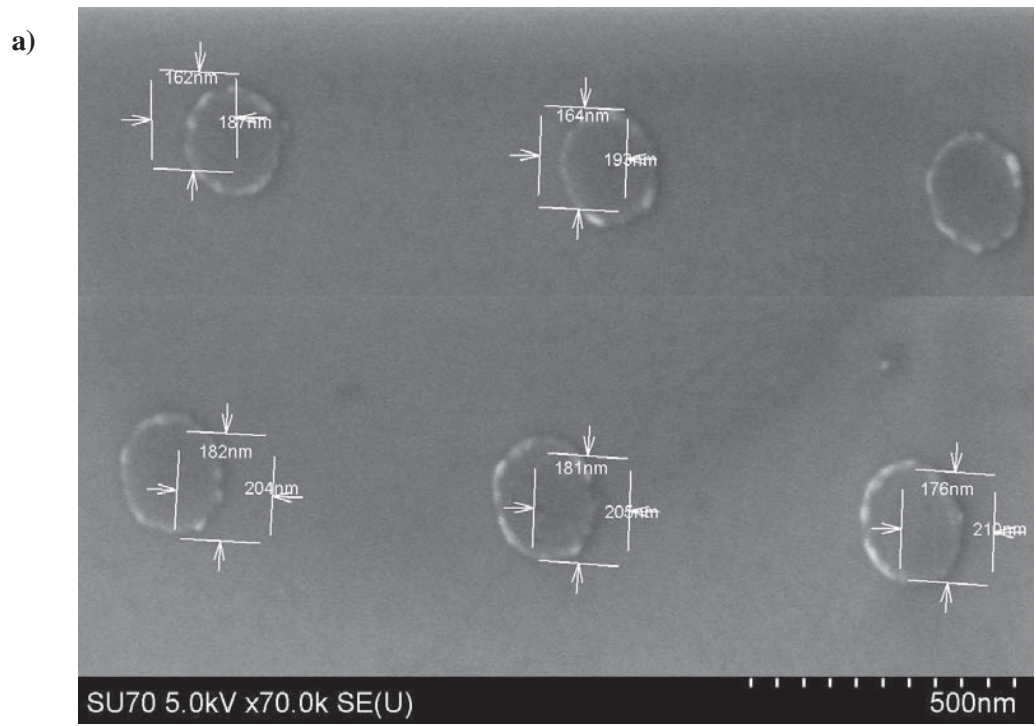

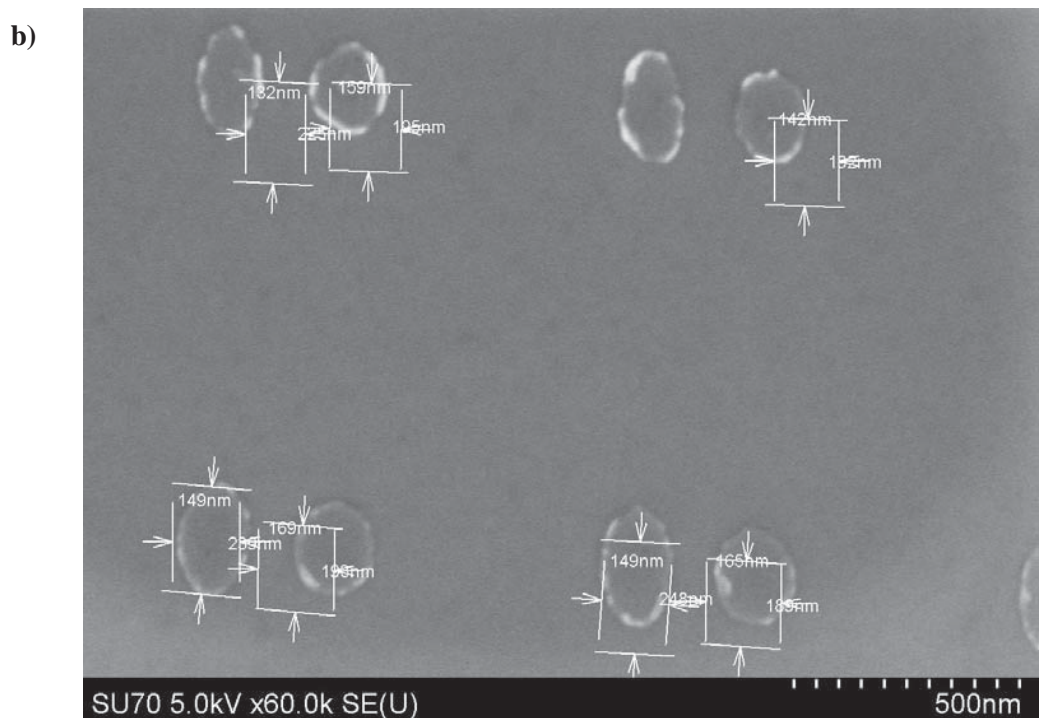

**Fig. S13. SEM images of a) isolated nanomagnet with nominal dimensions of (200 nm × 185 nm × 12 nm), and b) dipole-coupled nanomagnets with nominal dimensions of (250 nm × 150 nm × 12 nm, 200 nm × 175 nm × 12 nm), showing fabrication variation in lateral dimensions. Note: A slight drift in the electron beam during scanning resulted in a shift in the measurement markers.**

Hence, the best "confirmative statistics" that we could obtain was 2 out of 2 switching events. Multiple samples were fabricated and comprehensively analyzed for switching behavior. Each of these 5 samples shows switching events, although the yield of nanomagnets that switch is consistently low.

Further, if we focus on a specific nanomagnet (isolated), or pair (dipole coupled NOT gate), or array of three (dipole coupled Bennett clocking) in these samples, we find the switching events to be as follows:

**Table 1-3: Switching events (best) in Co nanomagnets on PMN-PT substrate after stress application**

| Nanomagnet dimensions | # of switching events/# of stress cycles (best results only across 5 different samples) |
|---|---|
| Isolated (high shape anisotropy) $250 \times 150 \times 12$ nm$^3$ | 0 out of 3 *(This particular nanomagnet is deliberately designed NOT to switch)* |
| Isolated (medium shape anisotropy) $200 \times 175 \times 12$ nm$^3$ | 1 out of 2 *(expect isolated magnets to only switch 50% of the time, as once stress is withdrawn, it could relax from the hard axis to either easy direction)* |
| Isolated (low shape anisotropy) $200 \times 185 \times 12$ nm$^3$ | 1 out of 2 OR 2 out of 3 OR 1 out of 3 *(For 3 cycles it cannot switch "1.5 times", it would switch 1 on 3 or 2 on 3 times)* |
| Dipole-coupled NOT gate ($250 \times 150 \times 12$ nm$^3$, $200 \times 175 \times 12$ nm$^3$) | 2 out of 2 *(only 2 cycles before substrate failure)* |
| Array of 3 magnets (Bennett clocking) ($250 \times 150 \times 12$ nm$^3$, $200 \times 175 \times 12$ nm$^3$, $200 \times 185 \times 12$ nm$^3$) | 1 out of 2 *(for 3 magnets, where 2 magnets can switch, the chance of getting "up", "down", "up" is 1 out of 4 or 25%. So, the fact this works 50% of the time shows dipole coupling affects the switching)* |

Table 1-3 highlights the best switching "statistics" that we encountered over *two-three* stress cycles across the 5 different PMN-PT substrates, for a particular nanomagnet (isolated, pair or array).

Experiments for each case - (a) isolated (b) dipole coupled pair (c) dipole coupled array - are shown below. As mentioned, these results are for specific nanomagnet(s) and represent the best possible switching "statistics" encountered over the maximum number of stress cycles possible prior to sample failure. They do not represent the 'yield' of the same sample, i.e. how many of the fabricated nanomagnets have dimensions conducive to switching. All the best case switching events reported in the table are shown below with MFM plots.

i) <u>Isolated Nanomagnets</u>

In the first scenario, isolated nanomagnets with 'high' shape anisotropy experience no switching events, as expected, due to the stress anisotropy energy being insufficient to overcome the shape anisotropy energy barrier of these nanomagnets. MFM results of one such set of magnets with high anisotropy across 3 cycles of stress (and the magnetization state prior to stress application) are shown in Fig. S14. Although there is no magnetization rotation due to the stress anisotropy being unable to overcome the very high shape anisotropy, we can see that by stress cycle 3, the magnetization gets weaker, possibly due to sample degradation after several weeks.

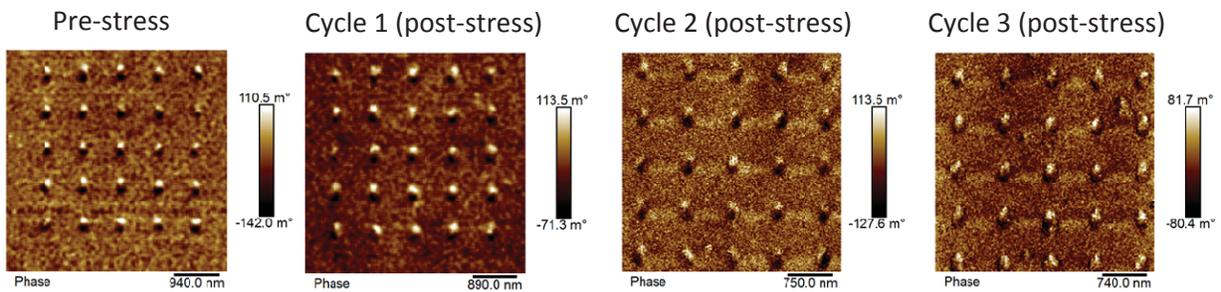

**Fig. S14. MFM images of high shape anisotropy nanomagnets (250×150×12 nm³) after 3 cycles of stress application, demonstrating that the stress in unable to induce magnetization rotation in these magnets.**

In nanomagnets with medium shape anisotropy, we noticed a maximum of 1 out of 2 switching events (Fig. S15), while magnets with low shape anisotropy showed 2 out of 3 switching events (Fig. S16). (For low anisotropy isolated nanomagnets 1 of 3 and 1 of 2 switching events were also observed and are not shown here). These are all consistent with what is expected; isolated nanomagnets would switch only 50% of the time.

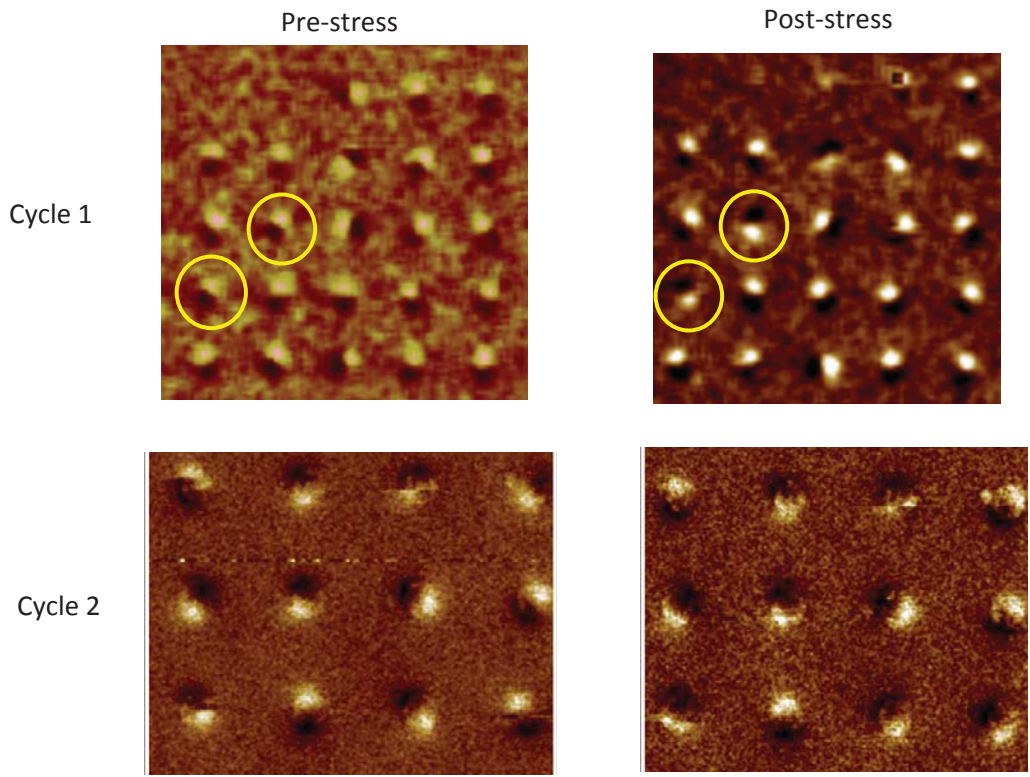

**Fig. S15. MFM images of lower shape anisotropy nanomagnets ($190 \times 185 \times 12$ nm$^3$) before and after stress.**

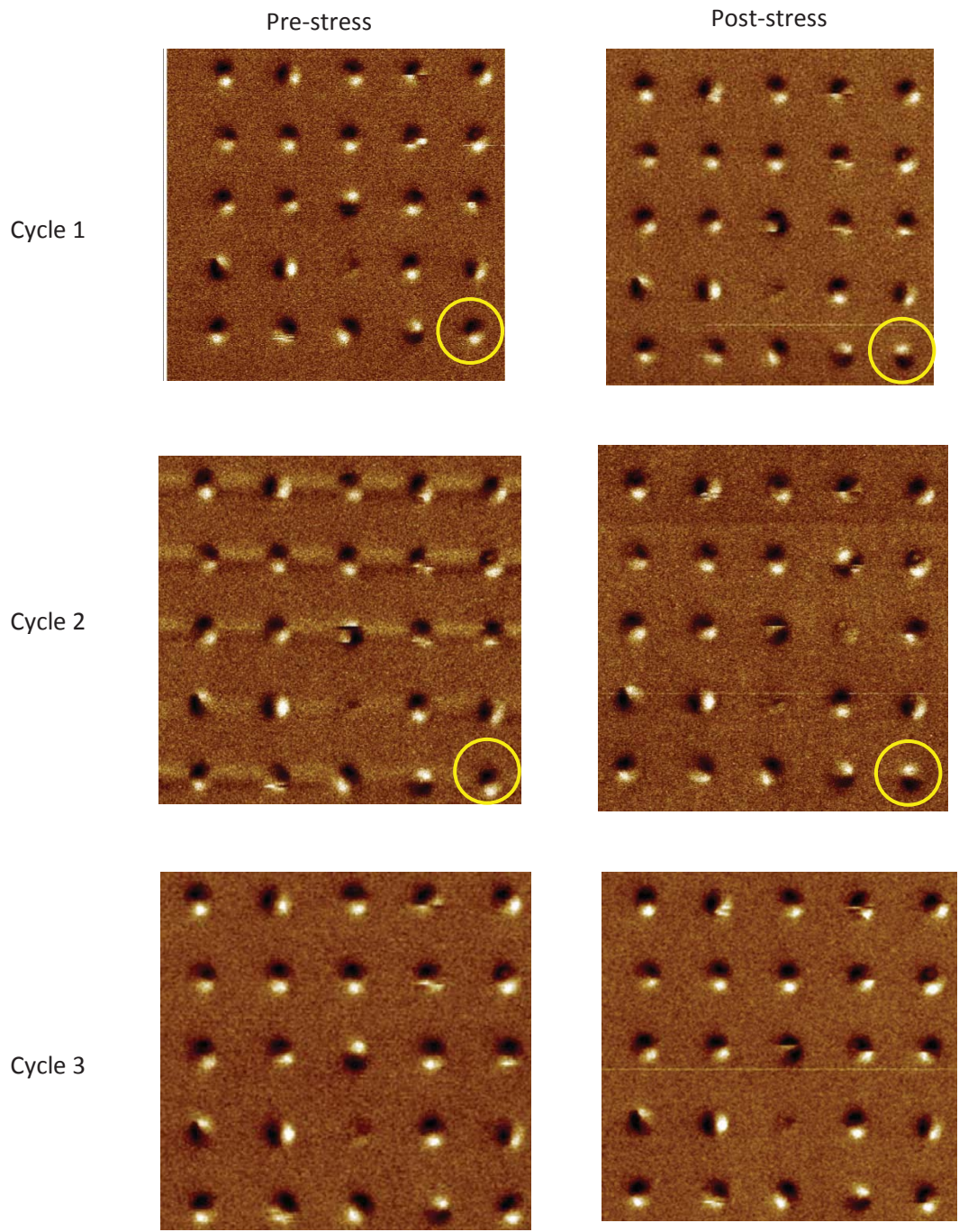

**Fig. S16. MFM images of dipole-coupled nanomagnets with lateral dimensions: (250 nm x 150 nm), (200 nm x 175 nm) for 3 stress cycles. Note: these nanomagnets are fabricated on a different PMN-PT substrate from that used in Fig. S15.**

Dipole-coupled NOT gate

In the case of dipole-coupled nanomagnets, magnetization switching due to the dipole interaction between the highly shape anisotropic nanomagnet (250 nm x 150 nm) and a lower shape anisotropy nanomagnet (200 nm x 175 nm) has been observed for the same magnet pair, as shown in Fig. S17.

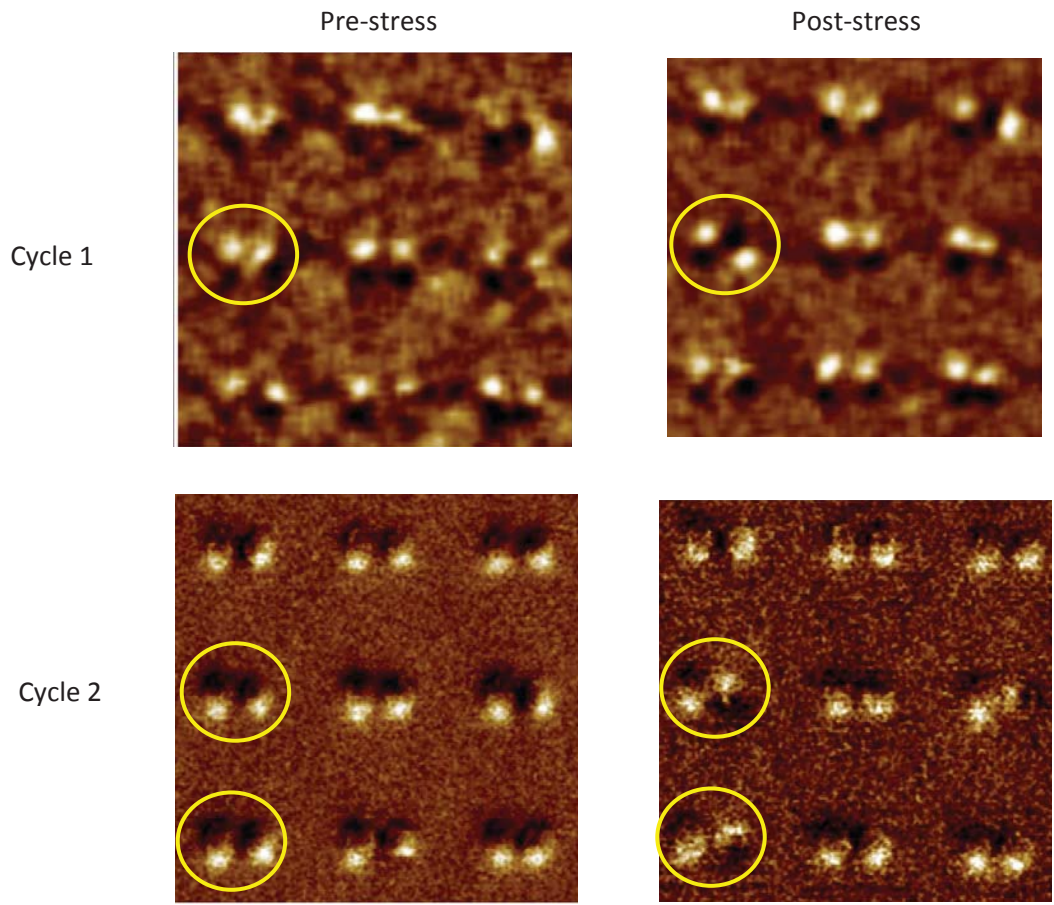

**Fig. S17. MFM images of dipole-coupled nanomagnets with lateral dimensions: (250 nm x 150 nm), (200 nm x 175 nm) for 2 stress cycles.**

Array of 3-dipole-coupled magnets (Bennett clocking)

Here, Bennett clocking was demonstrated with an array of 3 dipole-coupled nanomagnets with decreasing shape anisotropy. The low yield in this case stems from the constraint that the final nanomagnet has to have lower shape anisotropy than the central nanomagnet, while also ensuring that its energy barrier is not low enough that random thermal fluctuations or the stray field of the MFM tip can cause magnetization rotation. In Fig. S18, magnetization rotation to the desired state is observed after stress is applied. However, this was not repeatable in Cycle 2. Further stress cycles could not be applied owing to substrate failure.

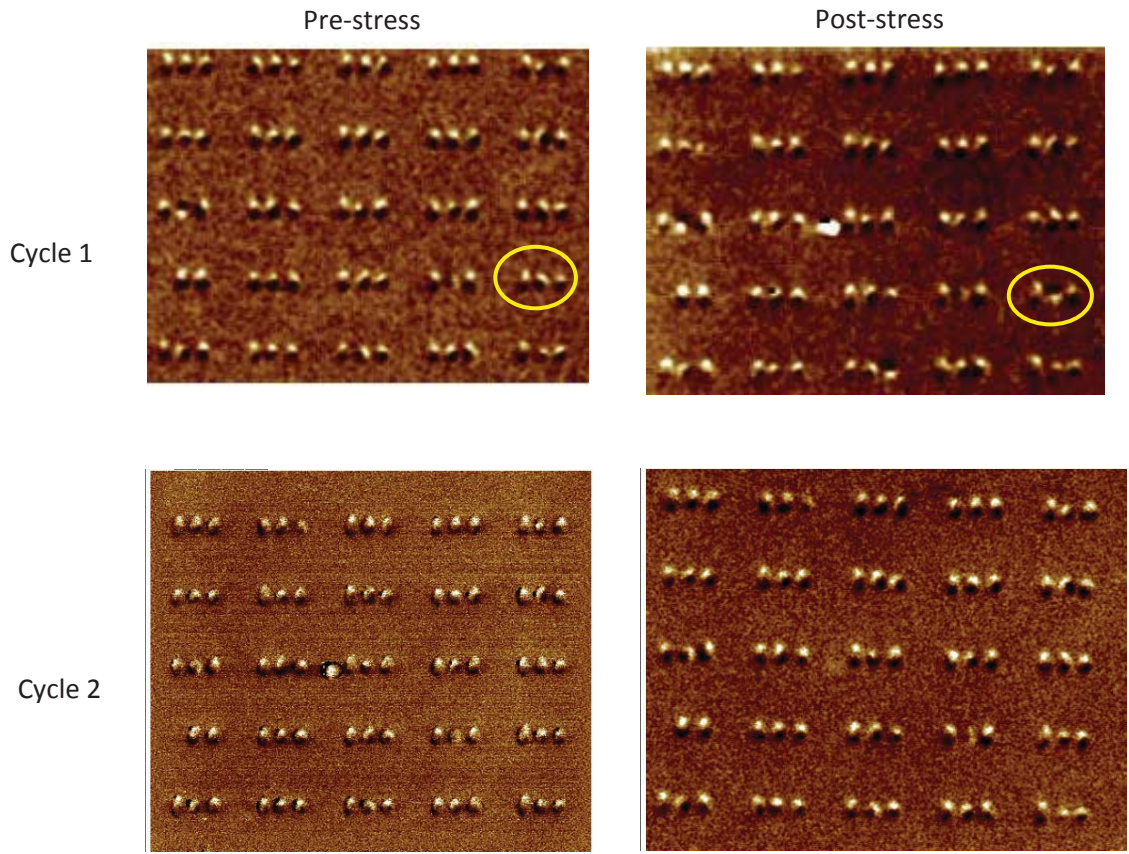

**Fig. S18. MFM images of 3-nanomagnet array with lateral dimensions: (250 nm x 150 nm), (200 nm x 175 nm), (200 nm x 185 nm) for 2 stress cycles.**

**Low Yield Explanation**

The low 'yield' of switching events in a set of fabricated nanomagnetic elements (isolated, pair or array) is one of the challenges of nanomagnetic logic (NML), where high switching error rates accruing from imperfect fabrication are frequently encountered. In recent work studying error rates in NML circuits with magnetic field-based switching (Shah et al., 2015), error rates as high as 77% were observed for low aspect ratio nanomagnetic elements and 76% for high aspect ratio elements in chains of nanomagnets. This was attributed to fabrication process-related variations, rough edges, etc. Only with careful fabrication methods (double e-beam exposure technique) that reduced the inter-magnet spacing, the error rates were brought down to 41% and 30% respectively, which are of course still very high.

**Switching "Statistics" Explanation**

Another important consideration when studying magnetization reversal due to stress in these magnetostrictive nanomagnets is that the effective magnetic field may just be able beat the shape anisotropy barrier but the switching can still be impeded by other effects such as pinning sites, defects, etc. The reason for the switching inconsistency in the nanomagnets (from stress cycle to stress cycle) can be explained as follows:

The stress anisotropy energy of a nanomagnet can be expressed as:

$$E_{stress-anisotropy} = \mu_0 M_s H_{eff} \sim \left(-\frac{3}{2}\lambda_s\right)\sigma$$

or, $$H_{eff} \sim \frac{1}{\mu_0 M_s}\left(-\frac{3}{2}\lambda_s\right)\sigma$$

where $H_{eff}$ is the effective magnetic field, $\mu_0$ is the permeability of free space ($4\pi \cdot 10^{-7}$ H·m$^{-1}$), $M_s$ is the saturation magnetization of Cobalt ($14.22 \times 10^5$ A·m$^{-1}$), ($\frac{3}{2}\lambda_s$) is the saturation magnetostriction of Co (50 ppm) and $\sigma$ is the stress applied to the nanomagnet (~ 80 MPa). All this yields a value of $H_{eff}$ ~ 30 Oe. While the nanomagnet's size and shape (low aspect ratio) are designed so that this "effective field due to stress" can beat the shape anisotropy barrier, the effective field (driving force) may not be sufficient to overcome the effects of pinning sites, edge roughness, etc, when considering the effects jagged edges, pinning sites, etc.

In fact, the M-H curve of Co film shows that the $H_{coercivity}$ is ~50 Oe (see the VSM data of the magnetization curves of films) but this is due to both substrate clamping and pinning sites. The $H_{coercivity}$ could be smaller in the nanostructures as the strain from the bulk substrate is transferred to the nanomagnet to switch it, so the substrate does not cause a "clamping effect". Hence, the coercivity in the magnets may be comparable to or smaller than $H_{eff}$ due to strain. As a result, some nanomagnets switch, but the "switching statistics" (fraction of successful switching events, even with maximum stress applied) is considerably low.

All this points to a fundamental limitation of strain clocking with Co nanomagnets. Clearly, if materials with better magnetoelastic coupling and higher magnetostriction could be fabricated (e.g. Terfenol-D, with 30 time higher magnetostriction), then the switching probability could improve significantly. This is because $H_{eff}$ due to strain ~ 1000 Oe for such materials stressed by ~80 MPa and can easily overcome any pinning sites and other defects that lead to high coercivity. Thus, the low yield of switching events in a large nanomagnet array could be attributed to the weak magnetoelastic coupling (material properties) of the magnetostrictive material investigated (Cobalt).

This process and materials-related issue, although highly important for technological applications, does not circumscribe the physics involved in strain-based magnetization switching. This has happened before, in other fields as well. Spin Hall effect was well known earlier, but only recent discover of large Spin-Hall angles in heavy metals (W and Ta) allowed efficient magnetization reversal using Spin-Hall effect. (Miron et al., 2011; Liu, et al.; Bhowmik, et al.)

## Supplementary Section D: Global vs. Local clocking and energy dissipation calculation

In the main paper, strain-induced magnetization switching in elliptical nanomagnets was accomplished using a *global* stress on a bulk PMN-PT substrate. A Boolean NOT logic gate was also demonstrated, along with unidirectional bit information propagation. An alternate scheme involves applying *local* stress in a phased manner to clock nanomagnets of nominally identical dimensions unidirectionally in the manner of reference 11 as shown in Fig. S19.

Each electrode pair is activated by applying an electrostatic potential between both members of that pair and the grounded substrate. Since the electrode in-plane dimensions are comparable to the piezoelectric film thickness, the out-of-plane ($d_{33}$) expansion/contraction and the in-plane ($d_{31}$) contraction/expansion of the piezoelectric regions underneath the electrodes produce a highly localized strain field under the electrodes[11]. Furthermore, since the electrodes are separated by a distance 1–2 times the piezoelectric film thickness, the interaction between the local strain fields below the electrodes will lead to a biaxial strain in the piezoelectric layer underneath the magnet[11]. This biaxial strain (compression/tension along the line joining the electrodes and tension/compression along the perpendicular axis) is transferred to the magnet, thus rotating its magnetization. This happens despite any substrate clamping and despite the fact that the electric field in the PZT layer just below the magnet is approximately zero since the metallic magnet shorts out the field[11]. The electrode pairs are activated sequentially in the manner shown in Fig. S19 to implement both NOT function and for unidirectional propagation of information along a chain of nanomagnets.

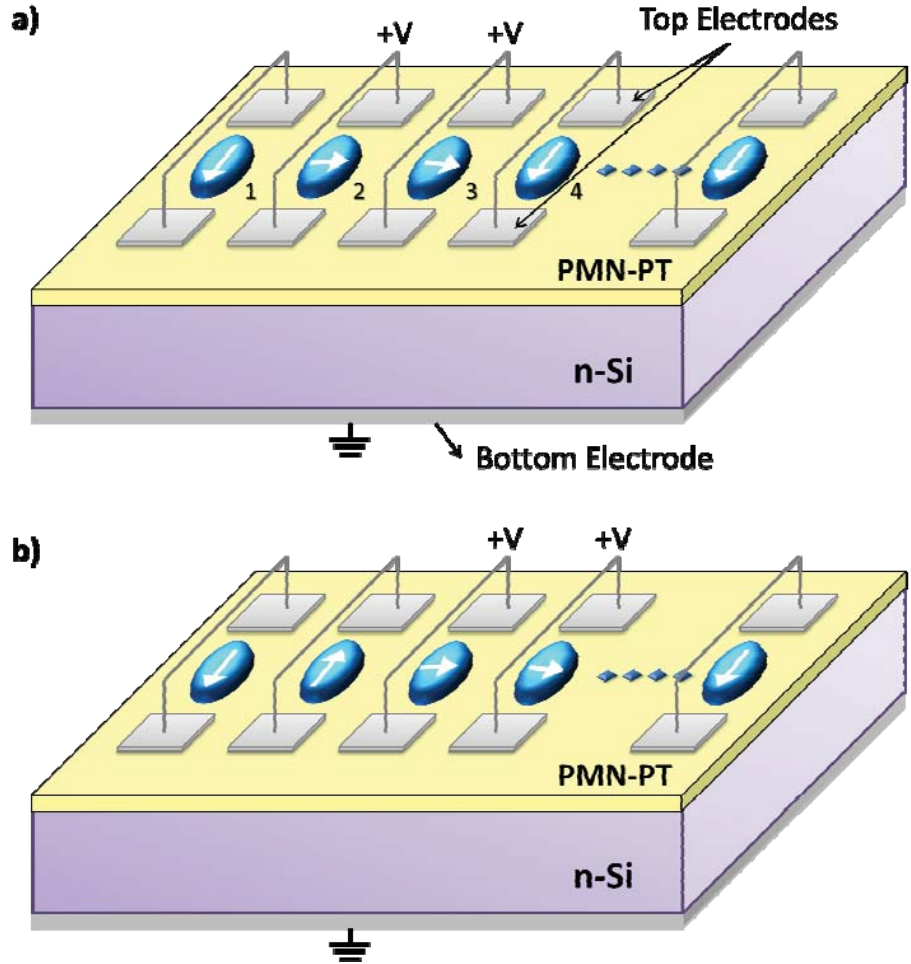

**Fig. S19: Local clocking of nanomagnets using the Bennett clocking scheme.** Ideally, if a local strain-clocking scheme is employed, stress can be applied *selectively* to targeted nanomagnets via individual electrodes[11]. Here all magnets are assumed to be nominally identical. (*a*) To propagate the magnetization state of the input magnet **1**, a voltage (+V) is applied to nanomagnets **2** and **3** simultaneously to generate a stress $\sigma$ to 'clock' them. (*b*) In the next phase of the clock cycle, the voltage (stress) is removed from **2**, while **3** and **4** are now clocked, resulting in the magnetization of **2** rotating and settling to the desired 'up' direction. This clock cycle is applied to successive nanomagnet pairs along the array with the input data propagating unidirectionally and replicated in every odd-numbered nanomagnet.

To highlight the potential energy efficiency of strain clocked nanomagnetic logic, we calculate the energy dissipation per clock cycle for the local clocking scheme. To generate a strain of ~400 ppm, a conservative estimate of the electric field needed for a PMN-PT film with $d_{33}$ = ~(1500-2500) pm/V [4] and $d_{31}$ = ~ -(700-1300) pm/V [4] in the above configuration is ~400 kV/m. To apply this field locally between the electrode and the substrate for a PMN-PT film of thickness t~200 nm, the voltage required would have been ~80 mV. The capacitance between the electrode pair and substrate is calculated by treating them as two flat plate capacitors in parallel. The area of each plate is A = $4 \times 10^{-14}$ m$^2$ (assume square electrode of width ~200 nm). The total capacitance including both electrodes is, C = $2 \varepsilon_0 \varepsilon_r A / t$ is ~10 fF. Assuming all the energy involved in charging the capacitor to strain the nanomagnet is lost, the energy dissipation/clock cycle, $E_d = 1/2 CV^2$ = $32 \times 10^{-18}$ J (32 aJ). Scaling the nanomagnet dimensions to ~100 nm and the square electrode width to ~100 nm will allow one to reduce the PMN-PT thickness to ~100 nm. This will reduce the switching voltage required to ~40 mV and the total capacitance to ~5 fF, making the energy dissipation go down to ~4 aJ. Moreover, if highly magnetostrictive materials such as Terfenol-D can be used instead of cobalt, the voltage needed can be decreased to ~8 mV and the energy dissipated in the switching circuit to ~0.16 aJ. Additional dissipation in the magnet due to Gilbert damping must then be taken into account and would roughly be ~1 aJ per clock cycle for a 1 GHz clock[12]. Therefore, the total dissipation in switching could be as low as ~1 aJ per clock cycle which is two to three orders of magnitude lower than what current transistors dissipate during switching[13] and one order of magnitude lower than the calculated dissipation in switching magnets with spin Hall effect[14]. That would make this scheme the most energy-efficient clocking mechanism extant.